\documentclass[10pt]{article}

\usepackage[utf8x]{inputenc}
\usepackage{graphicx} 
\usepackage{epsf}
\usepackage{a4}
\usepackage{booktabs}
\usepackage{caption}
\usepackage{textgreek}
\usepackage{tabularx}
\usepackage{amsmath}
\usepackage{amssymb}
\usepackage{authblk}
\usepackage{xcolor}
\usepackage{tikz}
\usepackage{color}
\usepackage{cite}
\usepackage{multirow}

\usetikzlibrary{shapes,arrows,snakes}

\usepackage[font=footnotesize,labelfont=bf]{caption}

\usepackage{geometry} 
\usepackage{graphicx} 

\usepackage{booktabs} 
\usepackage{array} 
\usepackage{paralist} 
\usepackage{verbatim} 
\usepackage{subfig} 

\usepackage{fancyhdr} 
\usepackage{sectsty}
\allsectionsfont{\sffamily\mdseries\upshape} 

\usepackage[nottoc,notlof,notlot]{tocbibind}
\usepackage[titles,subfigure]{tocloft} 
\usepackage{amsmath}
\usepackage{relsize}

\def\nn{\nonumber}
\def\beq{\begin{equation}}
\def\eeq{\end{equation}}
\def\bea{\begin{eqnarray}}
\def\eea{\end{eqnarray}}
\def\EQ{\begin{equation}}
\def\EN{\end{equation}}

\begin{document}

\title{Metastability in the Potts model: \\
exact results in the large $q$ limit}
\author{Onofrio Mazzarisi$^{1,2}$, Federico Corberi$^1$,  \\
Leticia F. Cugliandolo$^{2,3}$ and Marco Picco$^2$}
\affil{$^1$\textit{Dipartimento di Fisica E.R.Caianiello and INFN, gruppo collegato di Salerno, Universit\`a di Salerno, via Giovanni Paolo II 132, 8408 Fisciano (SA), Italy}}
\affil{$^2$\textit{Sorbonne Universit\'e, CNRS UMR 7589, Laboratoire de Physique Th\'eorique et Hautes Energies, 
4 Place Jussieu, 75252 Paris Cedex 05, France}}
\affil{$^3$\textit{Institut Universitaire de France, 1, rue Descartes, 75231 Paris Cedex 05, France}}

\maketitle
\thispagestyle{empty} 

\abstract{
We study the metastable equilibrium properties of the two dimensional Potts model with heat-bath transition rates
using a novel expansion. The method is especially powerful for large number of state spin variables
and it is notably accurate in a rather wide range of temperatures 
around the phase transition.
}

\newpage

\tableofcontents
\thispagestyle{empty} 

 \newpage
\setcounter{page}{1}

\section{Introduction}

The Potts model~\cite{Potts52} is an extension of the celebrated ferromagnetic Ising model. 
In this variation,  the spin variables take $q$ integer values (often associated to 
colours) and are coupled in a way that favours alignment, that is to say, equal values of the spins 
(colours) placed on neighbouring sites on a lattice. 
The model 
attracted attention at the early ages of phase transition studies since the order 
of the phase transition
changes when  the number of states of the spins is tuned: 
in two dimensions,
for $2 \le q \le 4$ it is of second-order, while for $q>4$ it is of first-order~\cite{Wu82,Baxter82}
with the associated metastability properties. 
Beyond the fundamental interest that 
it produced, the Potts model found applications in many areas of physics, 
and even beyond the physical domain. For instance, 
the large $q$  limit is used to describe soap foams and metallic grain systems~\cite{Weaire84,Stavans93,Glazier90}. 
In its anti-ferromagnetic version, the Potts model represents the
colouring problem of computer science~\cite{Sokal00,Salas01}. Another application in this realm
is to community detection in complex networks~\cite{Blatt96,Reichardt04,Ronhovde12}. 
Furthermore, 
weakly disordered Potts ferromagnets are the paradigmatic models in which the effects of randomness 
on phase transitions were studied~\cite{Dotsenko95a,Dotsenko95b}, 
and disordered and frustrated mean-field Potts models~\cite{KiTh88,ThKi88} realise the 
random first-order phase transitions 
scenario for the glassy arrest~\cite{KiThWo89,Biroli11,KiTh15}. 

The first order transition of the ferromagnetic two dimensional Potts model with $q>4$ is accompanied by metastability properties 
(with finite life-time in finite dimensions). In general, quantifying metastability 
and the dynamic escape from it through nucleation
is a hard and longstanding problem~\cite{Gunton83,Binder87,Oxtoby92,Kelton10}.
In this paper we address metastability in the stochastic bidimensional Potts model with $q>4$
from a novel perspective, that is, by solving the microscopic dynamics in the large $q$ limit. 
Indeed, in the stochastic model the dynamic evolution proceeds via a Markov Chain with microscopic 
rules that we have the freedom to choose, conditioned to respect detailed balance. As we argue below, 
the dynamics are faster, and also easier
to understand analytically, when the heat bath microscopic updates are used. This is the rule that we adopt.
The choice of initial conditions and working temperature decides the 
kind of metastability one accesses with the dynamic protocol. 
More precisely, for sub-critical quenches, in which we follow the evolution of 
a disordered initial state under conditions in which the system should order ferromagnetically, 
the metastable state is disordered. Instead, in the opposite quench, in which we prepare the system in a ferromagnetic state and we heat it above the 
critical point, the metastable state is 
ferromagnetically ordered. In this paper we consider both kinds of instantaneous 
quenches.  
After identifying the (few) relevant microscopic transition paths in the large $q$ limit, 
we derive the free-energy densities of the two phases and from them various  
thermodynamic observables that allow us to quantify the metastable behaviour 
in full detail. We confirm our analytical predictions with numerical simulations of excellent 
accuracy. 

The paper is organised as follows. In Sec.~\ref{sec:model} we recall the definition and main properties of the 
Potts model. In Sec.~\ref{sec:heat-bath} we introduce the heat bath dynamics, we identify all relevant moves 
for  $q>4$, and we derive the transition probabilities in terms of local configurations updates. Next, Sec.~\ref{sec:subcritical} and 
Sec.~\ref{sec:supercritical} describe our results for subcritical and supercritical quenches, respectively. A concluding Section
closes our work.

\section{The model}
\label{sec:model}

The Potts model~\cite{Potts52} is defined by the energy function
\begin{equation}
H_J[\{s_i\}] = - J \sum_{\langle ij \rangle} \delta_{s_is_j}
\; ,
\end{equation}
where $J>0$ is a coupling constant, the sum is restricted to nearest-neighbours on a lattice, 
$\delta_{ab}$ is the Kronecker delta and $s_i$ take integer values from 1 to $q \ge 2$. This model is a 
generalisation of the Ising model, to which it reduces for $q=2$. There is no external field applied. 
We will focus on the bidimensional case, defined on an $L \times L$ square lattice with periodic boundary conditions.
In the sum one counts each bond once and for this geometry  the energy is 
bounded between $-2JN$, with $N$ the number of spins in the sample, and $0$.

Although the problem is not fully solvable for $q>2$, some exact results are known. Duality allows one to prove that 
the critical temperature is~\cite{Potts52} 
\begin{equation}
k_BT_c (q) = \frac{J}{\ln\left(1+\sqrt{q} \right)}
\; . 
\end{equation} 
Henceforth we will set $k_B=J=1$.

An exact solution on the square lattice was provided in 1973:
by exploiting a mapping to the ice-rule six-vertex model
R. J. Baxter gave an exact expression for the model's 
free-energy {\it at the critical point}. He thus 
showed that the transition is second order for $q\leq 4$
and first order for $q>4$, and he calculated the latent heat in the latter case~\cite{Baxter73}.
A proof that the simplest possible mean-field approach yields, in the thermodynamic limit, the exact free-energy
at criticality for $q\geq q_c(d)$ (with $q_c(2)=4$) to leading order in $q$, in the large $q$ limit,
was soon after given by Mittal \& Stephen~\cite{Mittag74}, see also~\cite{Baracca83}. 
Many numerical studies put these ideas to the test since then. 
For example, Binder in Ref.~\cite{Binder81} and much more recently the authors of 
Refs.~\cite{Nam08,Huang10,Li18,Iino19} focused on the 
analysis of the critical properties, both in the second order and first order cases, 
using different numerical methods.

In order to go beyond the critical point results, 
F. Y. Wu exploited a fancy mapping onto a pure math problem to derive the free-energy density in the 
large $q$ limit at any $T$ (assuming that large $q$ and large $N$ limits commute)~\cite{Wu97} 
and he recovered the already known form at $T_c$~\cite{Baxter73,Mittag74} as a particular case. 
More recently, Johansson and Pistol used a microcanonical approach to argue that the entropy per 
site is given by~\cite{Johansson11}
\begin{equation}
s(e) = \left( 1 + \frac{e}{2} \right) \ln q
\end{equation}
with $e$ the energy density, 
in the large $N$ and $q$ limits, irrespectively of the order in which these are taken. 
They then used this result to calculate the 
partition function and from it the free-energy density
\begin{eqnarray}
-\beta f \sim
\left\{
\begin{array}{l}
\ln q \qquad 
\\
2\beta \qquad 
\end{array}
\right. 
\qquad
\mbox{for}
\qquad
\left\{
\begin{array}{l}
 \beta \ll \beta_c 
\\
 \beta \gg \beta_c
\end{array}
\right. 
\qquad
\mbox{with}
\quad 
\beta_c \simeq \ln \sqrt{q}
\end{eqnarray}
(in the last expression  $- \beta f \simeq -\beta e  \simeq 2\beta $ for large $\beta$ was used) 
that coincides with the one found in~\cite{Wu97}).

\section{Heat bath dynamics}
\label{sec:heat-bath}

Classical spin models coupled to heat baths evolve in time stochastically according to some 
microscopic updates that have to be provided to make their definition complete.
Concretely,  at each microscopic time step ones chooses one site at random 
and changes the value of the local spin according to some probabilistic rule. 
For a system with $N$ spins, conventionally, $N$ update attempts 
correspond to one Monte Carlo  step (MCs).  In this Section we define the Heat Bath microscopic rule, we enumerate
all possible updates of a chosen spin according to its surrounding configurations, and we derive the 
transition probability for each of them.

\subsection{Microscopic rules}

The usual microscopic dynamics used in Monte Carlo simulations of spin models are the Metropolis ones, 
in which one tries to change the spin to a new value (chosen at random among  
the remaining $q-1$ possibilities) and the move i)  is accepted if the new 
local energy $e_i'$ is lower than the previous local energy $e_i$ or, otherwise, ii) it is accepted 
with probability $\exp(-\beta (e'_i-e_i))$. 

However, in the case of the Potts model, especially in 
its large $q$ limit, another rule also respecting detailed balance, the so-called
{\it heat bath} rule, is more efficient and allows for a partial analytic treatment,
similarly to what found in other ferromagnetic models~\cite{bur09}.
In short, with this rule the transition probabilities are proportional to $e^{-\beta e'}$.
Specifically, the scheme  
works as follows. First, one considers the weight associated to each possible value that a spin, say $s_i$, 
can take depending on its local environment. 
As an example, assume that $s_i$  is surrounded, on the square lattice, 
by two spins taking the value $1$, a spin with value $2$ and another 
one with value $3$. We attribute the weights $w_i(s_i=1)= e^{2 \beta}$, 
corresponding to the fact that the spin $i$ taking the value $1$ yields a local energy of $-2$, 
$w_i(s_i=2)= e^{\beta} = w_i(s_i=3)$ because of the local energy being equal to $-1$ in these cases, 
and $w_i(s_i=j)=1$ for $ 3 < j \leq q$ for similar reasons. Next, we normalize the $w_i$ and we define 
the probabilities
\beq
P_i(s_i=k) = \frac{w_i(s_i=k)}{\sum_{l=1}^{q} w_i(s_i=l)}\; .
\eeq

Having attributed probabilities to the state of the central spin, we can now evaluate the transition probabilities for its
update. Imagine that the spin $s_i$ takes the value $1$. Then, we choose a random number $r \in [0:1]$.
If $r < P_i(1)$, the spin keeps its value $s_i=1$. Otherwise, if $r < P_i(1) + P_i(2)$, $s_i$ takes the new value 
$s_i=2$, or 
if $r < P_i(1) + P_i(2) + P_i(3)$, it is updated to $s_i = 3$, and so on and so forth. 
Thus, we have the following transition probabilities for the spin $s_i=1$ surrounded by two spins $1$, one spin $2$ and one spin $3$:
\begin{align}
& 
T^{\rm HB}_{1\to 1} = \frac{e^{2 \beta}}{e^{2 \beta} + 2 e^{ \beta} + q-3} \; , 
\qquad T^{\rm HB}_{1\to 2} = T^{\rm HB}_{1\to 3} = \frac{e^{\beta}}{e^{2 \beta} + 2 e^{ \beta} + q-3} \; , 
\\
& 
T^{\rm HB}_{1\to j} = \frac{1}{e^{2 \beta} + 2 e^{ \beta} + q-3}
\; , 
\end{align}
with $j$ indicating any possible state with $j>3$ (there are $q-3$ such states). Notice that these 
probabilities do not depend on the initial state of the spin. Despite this, we prefer to use the notation
above to make the comparison with the Metropolis probabilities (Eq.~(\ref{metropolis})). Proceeding in a 
similar way one can evaluate the transition probability of any spin, according to its 
state and the ones of its neighbours.

For the sake comparison, we recall the transition probabilities of the Metropolis rule:
\begin{align}
& 
T^{\rm M}_{1\to 1} = 1-\frac{1}{q-1}(2 e^{- \beta} + (q-3)e^{-2 \beta}) \; , \nonumber
\qquad 
T^{\rm M}_{1 \to 2} = T^{\rm M}_{1\to 3} = \frac{1}{q-1}e^{-\beta} 
\; , 
\\
&
T^{\rm M}_{1\to j} = \frac{1}{q-1}e^{-2 \beta} 
\; ,
\label{metropolis}
\end{align}
for the same example considered above.

In practice, we find that the heat-bath dynamics are much more efficient, in the sense that the 
approach to equilibrium is faster,  in particular for large $q$. 
We only consider the heat-bath dynamics in the following.

\subsection{Enumeration}

For any integer $q\geq 5$ we can classify all local configurations, seen as vertices with a central spin and 
its four first neighbours, and identify all possible updates. The method goes like this. 
Take one spin $s_i$,
count the number of neighbouring spins with the same value as the selected central one, 
and call this number $n_1$. 
Next, count the number of neighbours with the most present spin value different from the central 
one and call this number $n_2$. Continue in this way and organise these numbers in 
decreasing order, that is, 
$n_1, \, n_2, \, n_3, \dots$. It is easy to see that, with this classification, there are only 11 local configurations
(we do not distinguish which are the neighbours that take the same or different values as the central
one) and they are represented in the figure below:
\\
\\
\\
\begin{tikzpicture}[scale=0.7]
\node[circle] at (-2,0) { \ (0) : };
\node[circle,minimum size=0.5cm,fill=gray] at (-1.,0.0) {};
\node[circle,minimum size=0.5cm,fill=gray] at (0.,1.0) {};
\node[circle,minimum size=0.5cm,draw=red,fill=gray] at (0.,0.0) {};
\node[circle,minimum size=0.5cm,fill=gray] at (0.,-1.0) {};
\node[circle,minimum size=0.5cm,fill=gray] at (1.,0.0) {};
\end{tikzpicture}
\quad
\begin{tikzpicture}[scale=0.7]
\node[circle] at (-2,0) {\ (1) : };
\node[circle,minimum size=0.5cm,fill=blue] at (-1.,0.0) {};
\node[circle,minimum size=0.5cm,fill=gray] at (0.,1.0) {};
\node[circle,minimum size=0.5cm,draw=red,fill=gray] at (0.,0.0) {};
\node[circle,minimum size=0.5cm,fill=gray] at (0.,-1.0) {};
\node[circle,minimum size=0.5cm,fill=gray] at (1.,0.0) {};
\end{tikzpicture}
\quad
\begin{tikzpicture}[scale=0.7]
\node[circle] at (-2,0) {\ (2) : };
\node[circle,minimum size=0.5cm,fill=blue] at (-1.,0.0) {};
\node[circle,minimum size=0.5cm,fill=gray] at (0.,1.0) {};
\node[circle,minimum size=0.5cm,draw=red,fill=gray] at (0.,0.0) {};
\node[circle,minimum size=0.5cm,fill=blue] at (0.,-1.0) {};
\node[circle,minimum size=0.5cm,fill=gray] at (1.,0.0) {};
\end{tikzpicture}
\quad
\begin{tikzpicture}[scale=0.7]
\node[circle] at (-2,0) {\ (3) : };
\node[circle,minimum size=0.5cm,fill=blue] at (-1.,0.0) {};
\node[circle,minimum size=0.5cm,fill=gray] at (0.,1.0) {};
\node[circle,minimum size=0.5cm,draw=red,fill=gray] at (0.,0.0) {};
\node[circle,minimum size=0.5cm,fill=gray] at (0.,-1.0) {};
\node[circle,minimum size=0.5cm,fill=red] at (1.,0.0) {};
\end{tikzpicture}
\\\\
\begin{tikzpicture}[scale=0.7]
\node[circle] at (-2,0) {\ (4) : };
\node[circle,minimum size=0.5cm,fill=blue] at (-1.,0.0) {};
\node[circle,minimum size=0.5cm,fill=gray] at (0.,1.0) {};
\node[circle,minimum size=0.5cm,draw=red,fill=gray] at (0.,0.0) {};
\node[circle,minimum size=0.5cm,fill=blue] at (0.,-1.0) {};
\node[circle,minimum size=0.5cm,fill=blue] at (1.,0.0) {};
\end{tikzpicture}
\quad
\begin{tikzpicture}[scale=0.7]
\node[circle] at (-2,0) {\ (5) : };
\node[circle,minimum size=0.5cm,fill=blue] at (-1.,0.0) {};
\node[circle,minimum size=0.5cm,fill=gray] at (0.,1.0) {};
\node[circle,minimum size=0.5cm,draw=red,fill=gray] at (0.,0.0) {};
\node[circle,minimum size=0.5cm,fill=blue] at (0.,-1.0) {};
\node[circle,minimum size=0.5cm,fill=red] at (1.,0.0) {};
\end{tikzpicture}
\quad
\begin{tikzpicture}[scale=0.7]
\node[circle] at (-2,0) {\ (6) : };
\node[circle,minimum size=0.5cm,fill=blue] at (-1.,0.0) {};
\node[circle,minimum size=0.5cm,fill=gray] at (0.,1.0) {};
\node[circle,minimum size=0.5cm,draw=red,fill=gray] at (0.,0.0) {};
\node[circle,minimum size=0.5cm,fill=green] at (0.,-1.0) {};
\node[circle,minimum size=0.5cm,fill=red] at (1.,0.0) {};
\end{tikzpicture}
\quad
\begin{tikzpicture}[scale=0.7]
\node[circle] at (-2,0) {\ (7) : };
\node[circle,minimum size=0.5cm,fill=blue] at (-1.,0.0) {};
\node[circle,minimum size=0.5cm,fill=blue] at (0.,1.0) {};
\node[circle,minimum size=0.5cm,draw=red,fill=gray] at (0.,0.0) {};
\node[circle,minimum size=0.5cm,fill=blue] at (0.,-1.0) {};
\node[circle,minimum size=0.5cm,fill=blue] at (1.,0.0) {};
\end{tikzpicture}
\\\\
\begin{tikzpicture}[scale=0.7]
\node[circle] at (-2,0) {\ (8) : };
\node[circle,minimum size=0.5cm,fill=blue] at (-1.,0.0) {};
\node[circle,minimum size=0.5cm,fill=blue] at (0.,1.0) {};
\node[circle,minimum size=0.5cm,draw=red,fill=gray] at (0.,0.0) {};
\node[circle,minimum size=0.5cm,fill=blue] at (0.,-1.0) {};
\node[circle,minimum size=0.5cm,fill=red] at (1.,0.0) {};
\end{tikzpicture}
\quad
\begin{tikzpicture}[scale=0.7]
\node[circle] at (-2,0) {\ (9) : };
\node[circle,minimum size=0.5cm,fill=blue] at (-1.,0.0) {};
\node[circle,minimum size=0.5cm,fill=blue] at (0.,1.0) {};
\node[circle,minimum size=0.5cm,draw=red,fill=gray] at (0.,0.0) {};
\node[circle,minimum size=0.5cm,fill=red] at (0.,-1.0) {};
\node[circle,minimum size=0.5cm,fill=red] at (1.,0.0) {};
\end{tikzpicture}
\quad
\begin{tikzpicture}[scale=0.7]
\node[circle] at (-2,0) {(10) : };
\node[circle,minimum size=0.5cm,fill=blue] at (-1.,0.0) {};
\node[circle,minimum size=0.5cm,fill=blue] at (0.,1.0) {};
\node[circle,minimum size=0.5cm,draw=red,fill=gray] at (0.,0.0) {};
\node[circle,minimum size=0.5cm,fill=green] at (0.,-1.0) {};
\node[circle,minimum size=0.5cm,fill=red] at (1.,0.0) {};
\end{tikzpicture}
\quad
\begin{tikzpicture}[scale=0.7]
\node[circle] at (-2,0) {(11) : };
\node[circle,minimum size=0.5cm,fill=blue] at (-1.,0.0) {};
\node[circle,minimum size=0.5cm,fill=green] at (0.,1.0) {};
\node[circle,minimum size=0.5cm,draw=red,fill=gray] at (0.,0.0) {};
\node[circle,minimum size=0.5cm,fill=yellow] at (0.,-1.0) {};
\node[circle,minimum size=0.5cm,fill=red] at (1.,0.0) {};
\end{tikzpicture}
\\
\\
In the following we will use the name ``sand'' to refer to the configurations (11) in which all 
sites take different values.
We now use a more detailed notation to identify each of these configurations
writing explicitly the number of neighbours of each kind, 
that is to say, using $[n_1, n_2, \dots]$ where only the values $n_i \neq 0$ are kept.
Proceeding in this way we have
\begin{eqnarray*}
(0) &:&  [4] \rightarrow  (0) \; , \; (7) \\
(1) &:&  [3,1] \rightarrow (1) \; , \; (4)  \; , \;  (8)  \\
(2) &:& [2,2] \rightarrow  (2) \; , \;  (2) \; , \;  (9) \\
(3) &:&  [2 , 1 , 1 ]    \rightarrow           (3)  \; , \;  (5)  \; , \;  (10) \\
(4) &:&  [1 , 3 ]       \rightarrow            (4)  \; , \;  (1)  \; , \;   (8) \\
(5)  &:&  [1 , 2 , 1 ]    \rightarrow           (5)  \; , \;  (3)  \; , \;  (10)  \\
(6)  &:&  [1 , 1 , 1 , 1 ]  \rightarrow        (6)  \; , \;  (11) \\
(7) &:&   [0 , 4 ]         \rightarrow           (7)  \; , \;  (0) \\
(8) &:&  [0 , 3 , 1 ]     \rightarrow          (8) \; , \;   (1) \; , \;  (4) \\
(9)  &:&  [ 0 , 2 , 2 ]        \rightarrow     (9)  \; , \;  (2) \\
(10) &:&  [ 0 , 2 , 1 , 1 ]  \rightarrow     (10) \; , \;  (3)\; , \; (5) \\
(11) &:&  [ 0 , 1 , 1 , 1 , 1 ] \rightarrow    (11)  \; , \;  (6)  
\end{eqnarray*}
where the right arrows and the values after them indicate the transitions generated by
the update of the central spin. For example, the first configuration, denoted by $(0)$,
can either keep the same value, 
thus the $(0)$ on the right, or take another value, thus the configuration $(7) :  [0,4]$. 
Again, this should be easy to grasp by looking at the sketch above.

\subsection{Transition probabilities}
\label{subsec:transition}

For each local situation, we can then read the rules for the heat-bath dynamics.
The local configuration $(0)$ remains the same with probability $\simeq e^{4\beta}$ and changes to 
any of the other $q-1$ possible values of the spin with 
probability $e^{0}=1$. Then, normalising the probabilities, we obtain
\beq
P_{0 \rightarrow 0} = \frac{e^{4\beta}}{e^{4\beta}+q-1}  \; , \qquad\qquad \; P_{0 \rightarrow 7} = \frac{q-1}{e^{4\beta}+q-1} 
\; . 
\eeq
In a similar way, we derive all other transition probabilities: 
\begin{eqnarray}
\displaystyle{
\begin{array}{lll} 
P_{1 \rightarrow 1} = 
\displaystyle{\frac{e^{3\beta}}{e^{3\beta}+e^{\beta}+q-2}}  \; , 
&
P_{1 \rightarrow 4} = \displaystyle{\frac{e^{\beta}}{e^{3\beta}+e^{\beta}+q-2} }
\; ,  
&
P_{1 \rightarrow 8} = \displaystyle{\frac{q-2}{e^{3\beta}+e^{\beta}+q-2}  }
\; , 
\vspace{0.15cm}
\nn \\
P_{2 \rightarrow 2} = \displaystyle{ \frac{2 e^{2\beta}}{2 e^{2\beta}+q-2}  } \; , 
&
P_{2 \rightarrow 9} = \displaystyle{ \frac{q-2}{2 e^{2\beta}+q-2} } \; ,
&
\vspace{0.15cm}
\nn \\
P_{3 \rightarrow 3} = \displaystyle{ \frac{e^{2\beta}}{e^{2\beta}+2 e^{\beta}+q-3} } \; , 
&
P_{3 \rightarrow 5} = \displaystyle{\frac{2  e^{\beta}}{e^{2\beta}+2 e^{\beta}+q-3} } \; , 
& 
P_{3 \rightarrow 10} =\displaystyle{ \frac{q-3}{e^{2\beta}+2 e^{\beta}+q-3} } \; , 
\vspace{0.15cm}
\nn  \\
P_{4 \rightarrow 4} = \displaystyle{\frac{e^{\beta}}{e^{\beta}+e^{3\beta}+q-2}  } \; , 
&
P_{4 \rightarrow 1} = \displaystyle{\frac{e^{3\beta}}{e^{\beta}+e^{3\beta}+q-2} } \; , 
& 
P_{4 \rightarrow 8} = \displaystyle{\frac{q-2}{e^{\beta}+e^{3\beta}+q-2} } \; , 
\vspace{0.15cm}
 \nn \\
P_{5 \rightarrow 5} = \displaystyle{\frac{2 e^{\beta}}{2 e^{\beta}+  e^{2\beta}+q-3} } \; , 
&
P_{5 \rightarrow 3} = \displaystyle{\frac{e^{2\beta}}{2 e^{\beta}+  e^{2\beta}+q-3}  }
\; , 
&
P_{5 \rightarrow 10} = \displaystyle{ \frac{q-3}{2 e^{\beta}+  e^{2\beta}+q-3}  } \; , 
\vspace{0.15cm}
\nn \\
P_{6 \rightarrow 6} = \displaystyle{ \frac{4 e^{\beta}}{4 e^{\beta}+q-4} } \; , 
&
P_{6 \rightarrow 11} = \displaystyle{\frac{q-4}{4 e^{\beta}+q-4}  }
\; , 
&
\vspace{0.15cm}
\nn \\
P_{7 \rightarrow 7} = \displaystyle{\frac{q-1}{e^{4\beta}+q-1} } \; ,
& 
P_{7 \rightarrow 0} = \displaystyle{\frac{e^{4\beta}}{e^{4\beta}+q-1} } \; ,
&
\vspace{0.15cm}
\nn \\
P_{8 \rightarrow 8} = \displaystyle{\frac{q-2}{e^{3\beta}+e^\beta+q-2} }  \; , 
&
P_{8 \rightarrow 1} = \displaystyle{\frac{e^{3\beta}}{e^{3\beta}+e^\beta+q-2} }
\; , 
&
P_{8 \rightarrow 4} = \displaystyle{\frac{e^\beta}{e^{3\beta}+e^\beta+q-2}}  
\; , 
\vspace{0.15cm}
\nn \\
P_{9 \rightarrow 9} = \displaystyle{\frac{q-2}{2 e^{2\beta}+q-2}  }
\; , 
&
 P_{9 \rightarrow 2} =\displaystyle{ \frac{2 e^{2\beta}}{2 e^{2\beta}+q-2}  }
 \; , 
&
\vspace{0.15cm}
\nn \\
P_{10 \rightarrow 10} = \displaystyle{\frac{q-3}{e^{2\beta}+ 2 e^\beta + q-3}  }
\; , 
&
P_{10 \rightarrow 3} = \displaystyle{\frac{e^{2\beta}}{e^{2\beta}+ 2 e^\beta + q-3}  }
\; , 
&
P_{10 \rightarrow 5} = \displaystyle{\frac{2 e^{ \beta}}{e^{2\beta}+ 2 e^\beta + q-3}  }
\; , 
\vspace{0.15cm}
\nn \\
P_{11 \rightarrow 11} = \displaystyle{\frac{q-4}{4 e^{\beta}+q-4} } \; , 
&
P_{11 \rightarrow 6} = \displaystyle{\frac{4 e^{\beta}}{4 e^{\beta}+q-4}  }
\; . 
&
\end{array}
}
\end{eqnarray}

Note that for any spin in the bulk, that does not feel the boundary if there exists one, these expressions are 
independent of the system size. Their large $q$ limit will be established below, when we will simultaneously
decide the temperature range studied that will itself also vary with $q$.

\section{Sub-critical quenches: the disordered metastable phase}
\label{sec:subcritical}

Let us focus now on the first dynamic protocol, a quench to a subcritical temperature $T < T_c(q)$ from a completely 
disordered state, {\it i.e.}, an equilibrium configuration at \(T\rightarrow\infty\).

\subsection{Large $q$ and large $N$ behaviour}

Consider a totally random configuration, a typical initial state at $t=0$. 
The number of sites in the configurations labeled (a), with $a=0, \dots, 11$ as in
the sketch above, are $N_a(0) = [(q-1)/q^4]  \, \tilde N_a(0) \, N$
with 
\begin{eqnarray}
\begin{array}{lll}
 \tilde N_0(0)=1/(q-1) \; , \quad & \tilde N_1(0)=4 \; , \quad  & \tilde N_2(0)= 6 \; ,  
\vspace{0.15cm}
 \\
\tilde N_3(0) = 6 (q-2)\; , \quad & \tilde N_4(0)=4 \; ,  \quad  & \tilde N_5(0)=12 (q-2) 
\; , 
\vspace{0.15cm}
 \\
\tilde N_6(0)= 4 (q-2) (q-3) \; , 
\quad & 
\tilde  N_7(0) =  1 \; ,  \quad & \tilde N_8(0) = 4 (q-2) \; , 
  \vspace{0.15cm}
\\
\tilde N_9(0)=3 (q-2)\; , \quad  &  \tilde N_{10}(0)= 6  (q-2) (q-3)\; , 
\quad &
\vspace{0.15cm}
\\
\tilde N_{11}(0) = (q-2) (q-3) (q-4) 
\; . 
\end{array}
\end{eqnarray}
For large $q$, the state (11) largely dominates the disordered configuration since
\begin{equation}
N_{11}(0) \simeq N(q-1) (q-2) (q-3) (q-4)/q^4 \simeq N \; .
\end{equation} 
The next configurations in the hierarchy are the (6) and (10) ones with 
\begin{equation}
N_6(0) \simeq 4N/q \; , \qquad\qquad N_{10}(0) \simeq 6N/q
\; .  
\end{equation}
All the other states appear with a much lower probability, reduced by at least another power of $q$.

In the large  \(q\) limit we can also write
\begin{equation}
\label{eq:beta_q_relation}
e^\beta=e^{\beta_cT_c/T}=e^{T_c/T\ln(1+ \sqrt q)}=(1+ \sqrt q)^{T_c/T}\simeq q^{T_c/(2T)} \ .
\end{equation}
Thus, during an update of the full lattice, the probability that a state (11) be replaced by a state (6) can be
expressed as
\begin{equation}
\label{R116}
P_{11 \rightarrow 6} = \frac{4 e^{\beta}}{4 e^{\beta}+q-4} \simeq \frac{4 q^{T_c/(2T)}}{4 q^{T_c/(2T)} + q}
=\frac{1}{1 + \frac{1}{4}q^{1-T_c/(2T)}} \ ,
\end{equation}
showing that the temperature \(T=T_c/2\) plays a special role.
Indeed,  for $q\gg 1$
\begin{equation}
\label{R116}
P_{11 \rightarrow 6} \to 1
\qquad
\mbox{at}
\qquad
T<T_c/2
\; ,
\end{equation}
{\it i.e.},  the state (11) is completely unstable and the system tends to reorganise really fast
at these low temperatures.
In the same large $q$ limit, at the cross-over temperature, 
\begin{equation}
P_{11 \rightarrow 6} \rightarrow 4/5
\quad
\mbox{and}
\quad
P_{6 \rightarrow 11} = 1-P_{11 \rightarrow 6} \rightarrow 1/5
\qquad
\mbox{at}
\qquad
T=T_c/2
\; ,
\end{equation}
meaning that the states labeled (11) are again unstable, even though in a weaker way. The system will still 
reorganise at $T_c/2$. Finally, 
\begin{equation}
P_{11 \rightarrow 6} \rightarrow 0 
\qquad
\mbox{at}
\qquad
T>T_c/2
\; ,
\end{equation}
and the system remains disordered in the large $q$ limit, in the full temperature interval $(T_c/2,T_c]$.

\begin{figure}[h!]
\begin{center}
\scalebox{.7}{\input{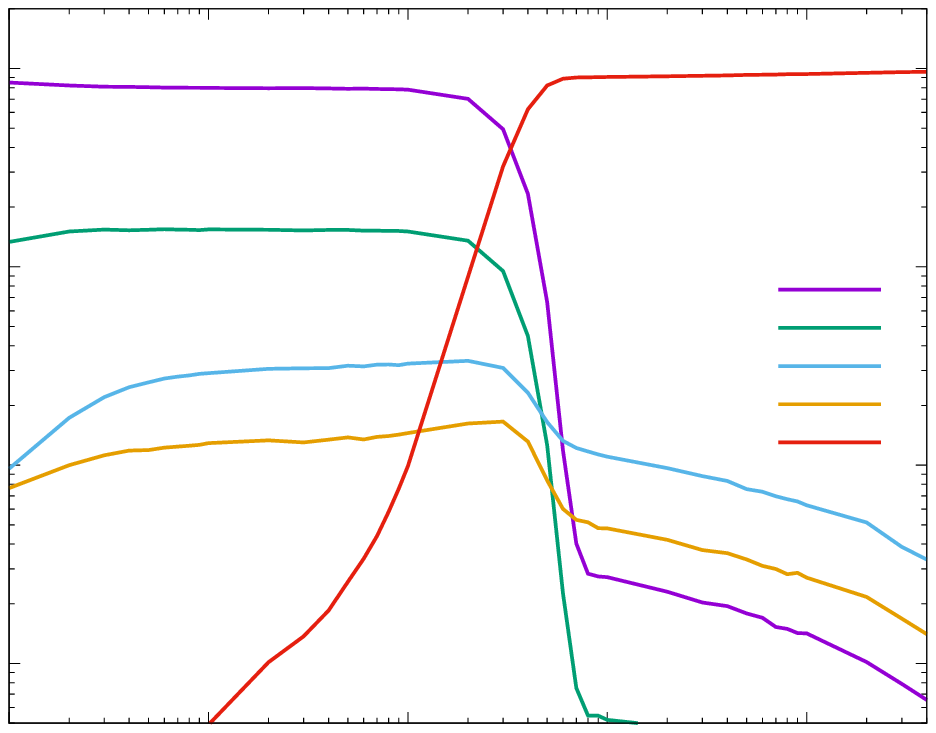}}
\hspace{1cm}
\end{center}
\caption{The time evolution of $N_a(t)/N$ for $a=0, 3, 6, 10, 11$ at $T=0.9 \, T_c$ in a square lattice system with linear size 
$L=10^3$ and $q=10^3$.}
\label{fig:NF1}
\end{figure}
%

\begin{figure}[h!]
\begin{center}
\includegraphics[scale=0.73]{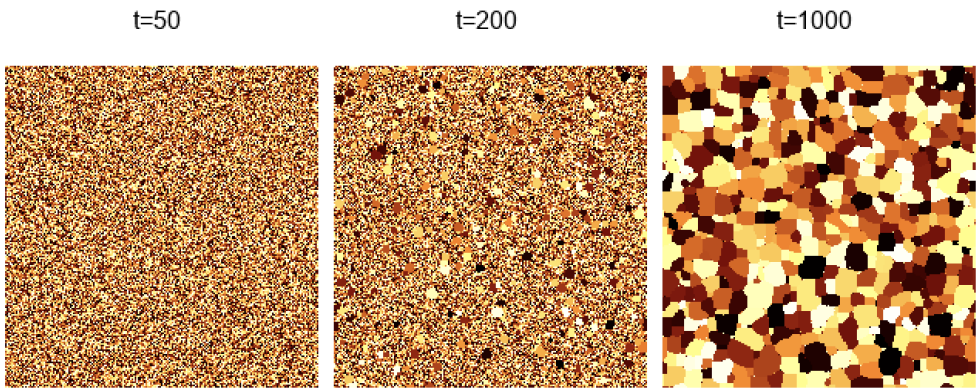}
\end{center}
\caption{Snapshots at times $t=50, 200, 1000$ for a square lattice system with linear size 
$L=10^3$ and $q=10^3$. Different colors are different spin values. 
}
\label{fig:NF2}
\end{figure}

When \(q\) is large but finite the picture is qualitatively similar, although the change is no longer at $T=T_c/2$ and 
it is not as sharp. The system does not in general remain disordered after a quench at \(T>T_c/2\) 
but it is only in this region that it can be found in a metastable state. To be more precise, let us consider 
a particular case. For a finite value of $q=10^3$ and after a quench at $T=0.9 \, T_c$, we observe the behaviour 
shown in Figs.~\ref{fig:NF1} and \ref{fig:NF2}. i) During a first period, most of the spins are in the (11) state and there are
only very small domains, the configurations look like sand.  The density of vertices (11) is almost 1, see Fig.~\ref{fig:NF1}, and 
the left snapshot in Fig.~\ref{fig:NF2} shows one such configuration. ii) At a later time, 
we see the appearance of the stable state (0) and some larger domains are formed, 
see the central snapshot in Fig.~\ref{fig:NF2}. For the chosen parameters $q$ and $T$, the crossover 
occurs at a time $t \simeq 100$. 
iii) At even  later times, most of the states are in the (0) state and large domains are formed,
see the right panel in Fig.~\ref{fig:NF2}. This is the proper coarsening regime.
Each of these three regimes is characterised by a different type of dynamical behaviour. We call them i) metastable,
ii) fast forming finite domains and iii) coarsening.

We found that the measurement of $N_a(t)/N$ is a very practical way of determining the type of dynamics. Next, 
we found that for a given value of $q$, the time $t$ at which the change of behaviour is observed depends strongly on the value of the temperature at which the system is quenched. In particular, if $T$ moves close to $T_c$, the system seems to be blocked in a metastable state forever. 
For $T=0.99 \, T_c$ and $q=10^3$, as we will see below, the system is not able to escape the metastable state. 

Thus for a given value of $q$, after a quench at $T < T_c$, we observe metastable states up to a time 
which seems to diverge at some temperature value that we parametrise as $r_t(q)  = T/T_c$.  
The quantity $r_t(q)$ does not seem to depend on the systems' linear 
size considered. We found numerically $r_t(q=10^3) \simeq 0.98$, $r_t(q=10^4) \simeq 0.94$, $r_t(q=10^5) \simeq 0.92$, $r_t(q=10^6) \simeq 0.90$ and $r_t(q=10^9) \simeq 0.87$.  Thus, as we increase $q$, the temperature above which we observe metastable states forever slowly decreases. Presumably, this quantity will go to $0.5$ 
in the limit of infinite $q$. 

For $T/T_c > r_t(q)$, we always observed metastable states. We will concentrate in the following in the study of these metastable states. 

\begin{figure}[h!]
\begin{center}
\scalebox{.7}{\input{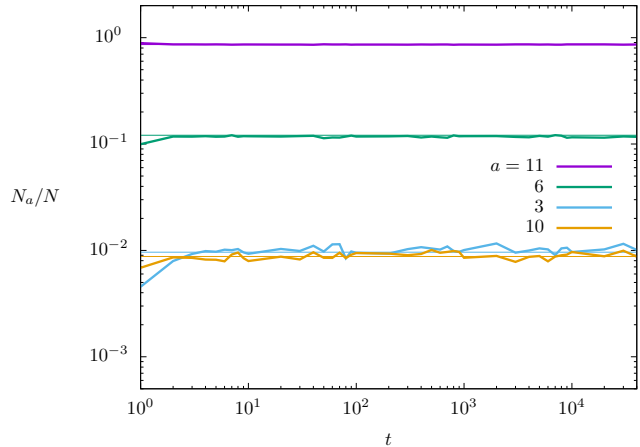}}
\hspace{1cm}
\end{center}
\caption{The time evolution of $N_a(t)/N$ for $a=3, 6, 10, 11$ at $T=0.99 \, T_c$ in a square lattice system with linear size 
$L=10^3$ and $q=10^3$. The thick lines are data from a numerical simulation while the 
thin ones are analytical predictions based on the method we develop in this work.
The curves demonstrate the hierarchy in Eq.~(\ref{hie}).}
\label{fig:F2}
\end{figure}

We illustrate the properties of these metastable states  in Fig.~\ref{fig:F2}, where we show the evolution of $N_a/N$ as a function 
of  time for $q=10^3$ and $L=10^3$ at $T/T_c =0.99$.  We only show the states which contribute the most.
Already at times of the order of $t\simeq 10^1$~MCs
after the quench, 
we found  $N_0(t) = N_2(t) = N_4(t) = N_7(t) = N_8(t) = N_9(t) =0$ 
while $0 \neq N_1(t) \simeq N_5(t) \simeq O(1) \ll N$ are not shown in the plot. The only values of order $N$ at
this time scale are $N_3$, $N_6$, $N_{10}$ and $N_{11}$. Their expected values,
according to the predictions based on the method we develop below, are 
$N_{11}/N \simeq 0.862, \, 
N_6/N \simeq 0.120, \, 
N_3/N \simeq  0.010, \, 
N_{10}/N \simeq  0.009$ 
and are shown with thin flat lines in the figure. The solid lines, instead, are the results of the numerical simulations, and are in 
excellent agreement with the analytic predictions.
Statistically, the configurations do not change after running the simulation much longer: the 
state made of ``vertices'' (3), (6), (10) and (11) according to the hierarchy 
\beq
\label{hie}
N_3(t) \simeq N_{10}(t) \ll N_6(t) \ll N_{11}(t) 
\eeq
with all of them being ${\mathcal O}(N)$, 
is metastable over incredibly long time-scales.

%

In the following, we concentrate on cases in which  $T$ is close to $T_c$. Moreover, we use 
the hierarchy relation (\ref{hie}) to develop an expansion that is notably accurate even keeping 
only the dominant order.

\subsection{The leading updates at $T \simeq T_c$}

We rename  \(N_a\) (\(a=0,...,11\)) the \textit{normalized} (by $N$) 
abundances that can also
be interpreted as the probabilities that a randomly picked site be in the state (a).
Exploiting the hierarchy relation~(\ref{hie}), 
expected to apply to the  metastable state,
we consider the evolution of 
\begin{equation}
N_{11} \simeq 1 \, , \qquad N_{6} \simeq p \, , \qquad N_{10} \simeq p^2
\qquad 
\mbox{and}
\qquad 
N_3 \simeq p^2
\end{equation}
thus rescaled with the parameter $p\equiv P_{11\rightarrow 6}$ which, at \(T\simeq T_c\),
is proportional to \(q^{-1/2}\):
\begin{equation}
p \equiv P_{11\rightarrow 6} \simeq q^{-1/2} \qquad\mbox{for} \qquad
T \sim T_c
\; .
\end{equation}
In the large $q$ limit, we will then use it as the small parameter in our expansion, that we will develop 
up to second order in  powers of $p$.

Concretely, our aim now is to construct a master equation for the probabilities $N_{11}$,  
$N_6$, $N_{10}$, $\dots$, 
and then find the stationary solution that determines 
the proportions of the vertices of each kind in the metastable states.

In order to do so, we first picture what kind of structures, {\it i.e.}, configurations of spins of the same color
(spin value) in a background of ``sand'' ({\it i.e.} spins in the (11) state)
have a probability to exist which is proportional to \(p^2\) or greater.
It turns out that spins in the states (6), (3) and (10), the only relevant ones in the large $N$ limit according 
to the discussion in the previous Subsection,
can only be found in the following structures

\begin{center}
\vspace{0.5cm}
\begin{tikzpicture}
\draw[thick][red]  (2.,0.) -- (3.,0.);
\node[circle,minimum size=0.8cm,draw=black,fill=white] at (0.,0.0) {11};
\node[circle,minimum size=0.8cm,draw=black,fill=white] at (1.,1.0) {11};
\node[circle,minimum size=0.8cm,draw=black,fill=white] at (1.,0.0) {11};
\node[circle,minimum size=0.8cm,draw=black,fill=white] at (1.,-1.0) {11};
\node[circle,minimum size=0.8cm,draw=black,fill=white] at (2.,2.0) {11};
\node[circle,label={[label distance=1cm]30:\textbf{A}},minimum size=0.8cm,draw=black,fill=white] at (2.,1.0) {11};
\node[circle,minimum size=0.8cm,draw=black,fill=lightgray] at (2.,0.0) {6};
\node[circle,minimum size=0.8cm,draw=black,fill=white] at (2.,-1.0) {11};
\node[circle,minimum size=0.8cm,draw=black,fill=white] at (2.,-2.0) {11};
\node[circle,minimum size=0.8cm,draw=black,fill=white] at (3.,1.0) {11};
\node[circle,minimum size=0.8cm,draw=black,fill=lightgray] at (3.,0.0) {6};
\node[circle,minimum size=0.8cm,draw=black,fill=white] at (3.,-1.0) {11};
\node[circle,minimum size=0.8cm,draw=black,fill=white] at (4.,0.0) {11};
\end{tikzpicture}
\qquad
\begin{tikzpicture}
\draw[thick][red]  (2.,0.) -- (3.,0.);
\draw[thick][red]  (1.0,0) -- (2.,0);
\node[circle,minimum size=0.8cm,draw=black,fill=white] at (0.,0.0) {11};
\node[circle,minimum size=0.8cm,draw=black,fill=white] at (1.,1.0) {11};
\node[circle,minimum size=0.8cm,draw=black,fill=lightgray] at (1.,0.0) {6};
\node[circle,minimum size=0.8cm,draw=black,fill=white] at (1.,-1.0) {11};
\node[circle,minimum size=0.8cm,draw=black,fill=white] at (2.,2.0) {11};
\node[circle,label={[label distance=1cm]30:\textbf{B}},minimum size=0.8cm,draw=black,fill=white] at (2.,1.0) {11};
\node[circle,minimum size=0.8cm,draw=black,fill=lightgray] at (2.,0.0) {3};
\node[circle,minimum size=0.8cm,draw=black,fill=white] at (2.,-1.0) {11};
\node[circle,minimum size=0.8cm,draw=black,fill=white] at (2.,-2.0) {11};
\node[circle,minimum size=0.8cm,draw=black,fill=white] at (3.,1.0) {11};
\node[circle,minimum size=0.8cm,draw=black,fill=lightgray] at (3.,0.0) {6};
\node[circle,minimum size=0.8cm,draw=black,fill=white] at (3.,-1.0) {11};
\node[circle,minimum size=0.8cm,draw=black,fill=white] at (4.,0.0) {11};
\end{tikzpicture}
\end{center}
\begin{center}
\begin{tikzpicture}
\draw[thick][red]  (2.,0.) -- (2.,1);
\draw[thick][red]  (2.,0) -- (3,0);
\node[circle,minimum size=0.8cm,draw=black,fill=white] at (0.,0.0) {11};
\node[circle,minimum size=0.8cm,draw=black,fill=white] at (1.,1.0) {11};
\node[circle,minimum size=0.8cm,draw=black,fill=white] at (1.,0.0) {11};
\node[circle,minimum size=0.8cm,draw=black,fill=white] at (1.,-1.0) {11};
\node[circle,minimum size=0.8cm,draw=black,fill=white] at (2.,2.0) {11};
\node[circle,label={[label distance=1cm]30:\textbf{C}},minimum size=0.8cm,draw=black,fill=lightgray] at (2.,1.0) {6};
\node[circle,minimum size=0.8cm,draw=black,fill=lightgray] at (2.,0.0) {3};
\node[circle,minimum size=0.8cm,draw=black,fill=white] at (2.,-1.0) {11};
\node[circle,minimum size=0.8cm,draw=black,fill=white] at (2.,-2.0) {11};
\node[circle,minimum size=0.8cm,draw=black,fill=white] at (3.,1.0) {10};
\node[circle,minimum size=0.8cm,draw=black,fill=lightgray] at (3.,0.0) {6};
\node[circle,minimum size=0.8cm,draw=black,fill=white] at (3.,-1.0) {11};
\node[circle,minimum size=0.8cm,draw=black,fill=white] at (4.,0.0) {11};
\end{tikzpicture}
\qquad
\begin{tikzpicture}
\draw[thick][red]  (2.,0.) -- (2.,1);
\draw[thick][red]  (2,0) -- (3.,0);
\draw[thick][red]  (3.,0.) -- (3.,1);
\draw[thick][red]  (2,1) -- (3.,1);
\node[circle,minimum size=0.8cm,draw=black,fill=white] at (0.,0.0) {11};
\node[circle,minimum size=0.8cm,draw=black,fill=white] at (1.,1.0) {11};
\node[circle,minimum size=0.8cm,draw=black,fill=white] at (1.,0.0) {11};
\node[circle,minimum size=0.8cm,draw=black,fill=white] at (1.,-1.0) {11};
\node[circle,minimum size=0.8cm,draw=black,fill=white] at (2.,2.0) {11};
\node[circle,label={[label distance=1cm]30:\textbf{D}},minimum size=0.8cm,draw=black,fill=lightgray] at (2.,1.0) {3};
\node[circle,minimum size=0.8cm,draw=black,fill=lightgray] at (2.,0.0) {3};
\node[circle,minimum size=0.8cm,draw=black,fill=white] at (2.,-1.0) {11};
\node[circle,minimum size=0.8cm,draw=black,fill=white] at (2.,-2.0) {11};
\node[circle,minimum size=0.8cm,draw=black,fill=lightgray] at (3.,1.0) {3};
\node[circle,minimum size=0.8cm,draw=black,fill=lightgray] at (3.,0.0) {3};
\node[circle,minimum size=0.8cm,draw=black,fill=white] at (3.,-1.0) {11};
\node[circle,minimum size=0.8cm,draw=black,fill=white] at (4.,0.0) {11};
\end{tikzpicture}
\end{center}
\begin{center}
\begin{tikzpicture}
\node[circle,minimum size=0.8cm,draw=black,fill=white] at (0.,0.0) {11};
\node[circle,minimum size=0.8cm,draw=black,fill=white] at (1.,1.0) {11};
\node[circle,minimum size=0.8cm,draw=black,fill=lightgray] at (1.,0.0) {11};
\node[circle,minimum size=0.8cm,draw=black,fill=white] at (1.,-1.0) {11};
\node[circle,minimum size=0.8cm,draw=black,fill=white] at (2.,2.0) {11};
\node[circle,label={[label distance=1cm]30:\textbf{E}},minimum size=0.8cm,draw=black,fill=white] at (2.,1.0) {11};
\node[circle,minimum size=0.8cm,draw=black,fill=white] at (2.,0.0) {10};
\node[circle,minimum size=0.8cm,draw=black,fill=white] at (2.,-1.0) {11};
\node[circle,minimum size=0.8cm,draw=black,fill=white] at (2.,-2.0) {11};
\node[circle,minimum size=0.8cm,draw=black,fill=white] at (3.,1.0) {11};
\node[circle,minimum size=0.8cm,draw=black,fill=lightgray] at (3.,0.0) {11};
\node[circle,minimum size=0.8cm,draw=black,fill=white] at (3.,-1.0) {11};
\node[circle,minimum size=0.8cm,draw=black,fill=white] at (4.,0.0) {11};
\end{tikzpicture}
\qquad
\begin{tikzpicture}
\node[circle,minimum size=0.8cm,draw=black,fill=white] at (0.,0.0) {11};
\node[circle,minimum size=0.8cm,draw=black,fill=white] at (1.,1.0) {11};
\node[circle,minimum size=0.8cm,draw=black,fill=white] at (1.,0.0) {11};
\node[circle,minimum size=0.8cm,draw=black,fill=white] at (1.,-1.0) {11};
\node[circle,minimum size=0.8cm,draw=black,fill=white] at (2.,2.0) {11};
\node[circle,label={[label distance=1cm]30:\textbf{F}},minimum size=0.8cm,draw=black,fill=lightgray] at (2.,1.0) {11};
\node[circle,minimum size=0.8cm,draw=black,fill=white] at (2.,0.0) {10};
\node[circle,minimum size=0.8cm,draw=black,fill=white] at (2.,-1.0) {11};
\node[circle,minimum size=0.8cm,draw=black,fill=white] at (2.,-2.0) {11};
\node[circle,minimum size=0.8cm,draw=black,fill=white] at (3.,1.0) {10};
\node[circle,minimum size=0.8cm,draw=black,fill=lightgray] at (3.,0.0) {11};
\node[circle,minimum size=0.8cm,draw=black,fill=white] at (3.,-1.0) {11};
\node[circle,minimum size=0.8cm,draw=black,fill=white] at (4.,0.0) {11};
\end{tikzpicture}
\vspace{0.25cm}
\end{center}

\noindent
where the gray sites in a given diagram possess the same color,
while the white sites have a different color with respect to the gray ones
and also with respect to the nearest and next-to-nearest other white ones.
The numbers indicate the kind of vertex, following the notation used
in the previous Subsections.
The red segments, which highlight the satisfied bonds, are useful to keep track of the energy 
contribution of the structures. It is possible
to check that all the other possible structures are of order \(p^3\) or higher
and we will not take them into account.

Now, we identify the evolutions that these structures can make in a single time
step. As an example consider structure \(\textbf B\). The following move 
\begin{center}
\vspace{0.25cm}
\begin{tikzpicture}
\draw[thick][red]  (1.,0.) -- (2.,0);
\draw[thick][red]  (2.0,0) -- (3,0);
\draw[thick][red]  (6.,0.) -- (7.,0);
\draw[thick][red]  (7.0,0) -- (8,0);
\draw[thick][red]  (8.,0.) -- (9.,0);
\node[circle,minimum size=0.8cm,draw=red,fill=white] at (0.,0.0) {11};
\node[circle,minimum size=0.8cm,draw=black,fill=white] at (1.,1.0) {11};
\node[circle,minimum size=0.8cm,draw=black,fill=lightgray] at (1.,0.0) {6};
\node[circle,minimum size=0.8cm,draw=black,fill=white] at (1.,-1.0) {11};
\node[circle,minimum size=0.8cm,draw=black,fill=white] at (2.,2.0) {11};
\node[circle,minimum size=0.8cm,draw=black,fill=white] at (2.,1.0) {11};
\node[circle,minimum size=0.8cm,draw=black,fill=lightgray] at (2.,0.0) {3};
\node[circle,minimum size=0.8cm,draw=black,fill=white] at (2.,-1.0) {11};
\node[circle,minimum size=0.8cm,draw=black,fill=white] at (2.,-2.0) {11};
\node[circle,minimum size=0.8cm,draw=black,fill=white] at (3.,1.0) {11};
\node[circle,minimum size=0.8cm,draw=black,fill=lightgray] at (3.,0.0) {6};
\node[circle,minimum size=0.8cm,draw=black,fill=white] at (3.,-1.0) {11};
\node[circle,minimum size=0.8cm,draw=black,fill=white] at (4.,0.0) {11};
\draw [very thick] [ ->] (4.6,0.0) -- (5.4,0.0);
\node[circle,minimum size=0.8cm,draw=red,fill=lightgray] at (6.,0.0) {6};
\node[circle,minimum size=0.8cm,draw=black,fill=white] at (7.,1.0) {11};
\node[circle,minimum size=0.8cm,draw=black,fill=lightgray] at (7.,0.0) {3};
\node[circle,minimum size=0.8cm,draw=black,fill=white] at (7.,-1.0) {11};
\node[circle,minimum size=0.8cm,draw=black,fill=white] at (8.,2.0) {11};
\node[circle,minimum size=0.8cm,draw=black,fill=white] at (8.,1.0) {11};
\node[circle,minimum size=0.8cm,draw=black,fill=lightgray] at (8.,0.0) {3};
\node[circle,minimum size=0.8cm,draw=black,fill=white] at (8.,-1.0) {11};
\node[circle,minimum size=0.8cm,draw=black,fill=white] at (8.,-2.0) {11};
\node[circle,minimum size=0.8cm,draw=black,fill=white] at (9.,1.0) {11};
\node[circle,minimum size=0.8cm,draw=black,fill=lightgray] at (9.,0.0) {6};
\node[circle,minimum size=0.8cm,draw=black,fill=white] at (9.,-1.0) {11};
\node[circle,minimum size=0.8cm,draw=black,fill=white] at (10.,0.0) {11};
\end{tikzpicture}
\vspace{0.25cm}
\end{center}
consists of a spin in state (11) turning into a state (6) and thus forming the structure on the right. 
The probability of this move is negligible
because the probability to pick a (11) which is around the structure on the left
(which contains (3)) is proportional to \(p^2\) and the probability now for it
to become a (6) is proportional to \(p\). The result is therefore proportional to $p^3$ and hence negligible at the 
order we are keeping. This kind of analysis can be performed 
for all the cases and thus prove that the structures labelled \(\textbf A\) to \(\textbf F\)
are at most of order \(p^2\) and every other is negligible.

The next step is to list all the possible moves that are relevant for the second order 
of our expansion and understand what are the consequences of each of these moves.
This will allow  us to
write down all the terms of the master equations for the probabilities \(N_{11}\),
\(N_{6}\), \(N_{3}\) and \(N_{10}\).
In practice we find that for (3) and (10) we need an equation for each of the configurations
in which these states can be found so we define the following quantities
\begin{center}
\vspace{0.25cm}
\begin{tikzpicture}
\draw[thick][red]  (2.,0.) -- (3.,0.);
\draw[thick][red]  (1.0,0) -- (2.,0);
\node[circle,minimum size=0.8cm,draw=black,fill=white] at (0.,0.0) {11};
\node[circle,minimum size=0.8cm,draw=black,fill=white] at (1.,1.0) {11};
\node[circle,minimum size=0.8cm,draw=black,fill=lightgray] at (1.,0.0) {6};
\node[circle,minimum size=0.8cm,draw=black,fill=white] at (1.,-1.0) {11};
\node[circle,minimum size=0.8cm,draw=black,fill=white] at (2.,2.0) {11};
\node[circle,label={[label distance=1cm]30:(3a)},minimum size=0.8cm,draw=black,fill=white] at (2.,1.0) {11};
\node[circle,minimum size=0.8cm,draw=red,fill=lightgray] at (2.,0.0) {3};
\node[circle,minimum size=0.8cm,draw=black,fill=white] at (2.,-1.0) {11};
\node[circle,minimum size=0.8cm,draw=black,fill=white] at (2.,-2.0) {11};
\node[circle,minimum size=0.8cm,draw=black,fill=white] at (3.,1.0) {11};
\node[circle,minimum size=0.8cm,draw=black,fill=lightgray] at (3.,0.0) {6};
\node[circle,minimum size=0.8cm,draw=black,fill=white] at (3.,-1.0) {11};
\node[circle,minimum size=0.8cm,draw=black,fill=white] at (4.,0.0) {11};
\end{tikzpicture}
\qquad
\begin{tikzpicture}
\draw[thick][red]  (2.,0.) -- (2.,1);
\draw[thick][red]  (2.,0) -- (3,0);
\node[circle,minimum size=0.8cm,draw=black,fill=white] at (0.,0.0) {11};
\node[circle,minimum size=0.8cm,draw=black,fill=white] at (1.,1.0) {11};
\node[circle,minimum size=0.8cm,draw=black,fill=white] at (1.,0.0) {11};
\node[circle,minimum size=0.8cm,draw=black,fill=white] at (1.,-1.0) {11};
\node[circle,minimum size=0.8cm,draw=black,fill=white] at (2.,2.0) {11};
\node[circle,label={[label distance=1cm]30:(3b)},minimum size=0.8cm,draw=black,fill=lightgray] at (2.,1.0) {6};
\node[circle,minimum size=0.8cm,draw=red,fill=lightgray] at (2.,0.0) {3};
\node[circle,minimum size=0.8cm,draw=black,fill=white] at (2.,-1.0) {11};
\node[circle,minimum size=0.8cm,draw=black,fill=white] at (2.,-2.0) {11};
\node[circle,minimum size=0.8cm,draw=black,fill=white] at (3.,1.0) {10};
\node[circle,minimum size=0.8cm,draw=black,fill=lightgray] at (3.,0.0) {6};
\node[circle,minimum size=0.8cm,draw=black,fill=white] at (3.,-1.0) {11};
\node[circle,minimum size=0.8cm,draw=black,fill=white] at (4.,0.0) {11};
\end{tikzpicture}
\qquad
\begin{tikzpicture}
\draw[thick][red]  (2.,0.) -- (2.,1);
\draw[thick][red]  (2,0) -- (3.,0);
\draw[thick][red]  (3.,0.) -- (3.,1);
\draw[thick][red]  (2,1) -- (3.,1);
\node[circle,minimum size=0.8cm,draw=black,fill=white] at (0.,0.0) {11};
\node[circle,minimum size=0.8cm,draw=black,fill=white] at (1.,1.0) {11};
\node[circle,minimum size=0.8cm,draw=black,fill=white] at (1.,0.0) {11};
\node[circle,minimum size=0.8cm,draw=black,fill=white] at (1.,-1.0) {11};
\node[circle,minimum size=0.8cm,draw=black,fill=white] at (2.,2.0) {11};
\node[circle,label={[label distance=1cm]30:(3c)},minimum size=0.8cm,draw=black,fill=lightgray] at (2.,1.0) {3};
\node[circle,minimum size=0.8cm,draw=red,fill=lightgray] at (2.,0.0) {3};
\node[circle,minimum size=0.8cm,draw=black,fill=white] at (2.,-1.0) {11};
\node[circle,minimum size=0.8cm,draw=black,fill=white] at (2.,-2.0) {11};
\node[circle,minimum size=0.8cm,draw=black,fill=lightgray] at (3.,1.0) {3};
\node[circle,minimum size=0.8cm,draw=black,fill=lightgray] at (3.,0.0) {3};
\node[circle,minimum size=0.8cm,draw=black,fill=white] at (3.,-1.0) {11};
\node[circle,minimum size=0.8cm,draw=black,fill=white] at (4.,0.0) {11};
\end{tikzpicture}
\vspace{0.25cm}
\end{center}
and 
\begin{center}
\vspace{1cm}
\begin{tikzpicture}
\node[circle,minimum size=0.8cm,draw=black,fill=white] at (0.,0.0) {11};
\node[circle,minimum size=0.8cm,draw=black,fill=white] at (1.,1.0) {11};
\node[circle,minimum size=0.8cm,draw=black,fill=lightgray] at (1.,0.0) {11};
\node[circle,minimum size=0.8cm,draw=black,fill=white] at (1.,-1.0) {11};
\node[circle,minimum size=0.8cm,draw=black,fill=white] at (2.,2.0) {11};
\node[circle,label={[label distance=1cm]30:(10a)},minimum size=0.8cm,draw=black,fill=white] at (2.,1.0) {11};
\node[circle,minimum size=0.8cm,draw=red,fill=white] at (2.,0.0) {10};
\node[circle,minimum size=0.8cm,draw=black,fill=white] at (2.,-1.0) {11};
\node[circle,minimum size=0.8cm,draw=black,fill=white] at (2.,-2.0) {11};
\node[circle,minimum size=0.8cm,draw=black,fill=white] at (3.,1.0) {11};
\node[circle,minimum size=0.8cm,draw=black,fill=lightgray] at (3.,0.0) {11};
\node[circle,minimum size=0.8cm,draw=black,fill=white] at (3.,-1.0) {11};
\node[circle,minimum size=0.8cm,draw=black,fill=white] at (4.,0.0) {11};
\end{tikzpicture}
\qquad
\begin{tikzpicture}
\node[circle,minimum size=0.8cm,draw=black,fill=white] at (0.,0.0) {11};
\node[circle,minimum size=0.8cm,draw=black,fill=white] at (1.,1.0) {11};
\node[circle,minimum size=0.8cm,draw=black,fill=white] at (1.,0.0) {11};
\node[circle,minimum size=0.8cm,draw=black,fill=white] at (1.,-1.0) {11};
\node[circle,minimum size=0.8cm,draw=black,fill=white] at (2.,2.0) {11};
\node[circle,label={[label distance=1cm]30:(10b)},minimum size=0.8cm,draw=black,fill=lightgray] at (2.,1.0) {11};
\node[circle,minimum size=0.8cm,draw=red,fill=white] at (2.,0.0) {10};
\node[circle,minimum size=0.8cm,draw=black,fill=white] at (2.,-1.0) {11};
\node[circle,minimum size=0.8cm,draw=black,fill=white] at (2.,-2.0) {11};
\node[circle,minimum size=0.8cm,draw=black,fill=white] at (3.,1.0) {10};
\node[circle,minimum size=0.8cm,draw=black,fill=lightgray] at (3.,0.0) {11};
\node[circle,minimum size=0.8cm,draw=black,fill=white] at (3.,-1.0) {11};
\node[circle,minimum size=0.8cm,draw=black,fill=white] at (4.,0.0) {11};
\end{tikzpicture}
\qquad
\begin{tikzpicture}
\draw[thick][red]  (3.,0.) -- (3.,1);
\draw[thick][red]  (2,1) -- (3.,1);
\node[circle,minimum size=0.8cm,draw=black,fill=white] at (0.,0.0) {11};
\node[circle,minimum size=0.8cm,draw=black,fill=white] at (1.,1.0) {11};
\node[circle,minimum size=0.8cm,draw=black,fill=white] at (1.,0.0) {11};
\node[circle,minimum size=0.8cm,draw=black,fill=white] at (1.,-1.0) {11};
\node[circle,minimum size=0.8cm,draw=black,fill=white] at (2.,2.0) {11};
\node[circle,label={[label distance=1cm]30:(10c)},minimum size=0.8cm,draw=black,fill=lightgray] at (2.,1.0) {3};
\node[circle,minimum size=0.8cm,draw=red,fill=white] at (2.,0.0) {10};
\node[circle,minimum size=0.8cm,draw=black,fill=white] at (2.,-1.0) {11};
\node[circle,minimum size=0.8cm,draw=black,fill=white] at (2.,-2.0) {11};
\node[circle,minimum size=0.8cm,draw=black,fill=lightgray] at (3.,1.0) {3};
\node[circle,minimum size=0.8cm,draw=black,fill=lightgray] at (3.,0.0) {3};
\node[circle,minimum size=0.8cm,draw=black,fill=white] at (3.,-1.0) {11};
\node[circle,minimum size=0.8cm,draw=black,fill=white] at (4.,0.0) {11};
\end{tikzpicture}
\vspace{2cm}
\end{center}

We can now express the probabilities for all the structures introduced above in terms of the 
probabilities of the various states
\begin{equation}
\begin{aligned}
P(\textbf A)&= (N_6-2N_{3a}-2N_{3b})/2 \; , \\
P(\textbf B)&= N_{3a} \; , \\
P(\textbf C)&= N_{3b} = N_{10c} \; , \\
P(\textbf D)&= N_{3c}/4 \; , \\
P(\textbf E)&= N_{10a} \; , \\
P(\textbf F)&= N_{10b}/2 \; , \\
\end{aligned}
\end{equation}
where the first one comes from the fact that for every two (6) which 
are not in the structure \(\textbf B\) or \(\textbf C\) (which contain two (6) each)
we count a structure \(\textbf A\). The derivation of $P(\textbf B), \dots, P(\textbf F)$ 
is straightforward. These expressions turn out to be useful to write down the probabilities 
of the moves, as we explain below.

Let us start with all the moves that a site which is in (11) can make.
Pick a site in (11) which is not a neighbor of any structure and turn it into a (6).
The probability for this move is 
\begin{equation}
P_{11\rightarrow6}=p ,
\end{equation} 
where we mean the extended, temperature and \(q\) dependent, form as in Eq. \ref{R116},
times the probability of picking such a 
(11) state. The latter equals \(N_{11}-3N_6\) because 
there are 3 sites in state \(N_{11}\) surrounding every 
(6) in structure \textbf A,  and we are neglecting the other terms of \(P(\textbf A)\)
and the other structures because they will lead to contributions of higher orders.
In this move we lose 2 (11) states and we gain 2 (6) states. In the following 
sketch we represent the move, we give its probability $P$ and we indicate below the 
sketch the loss and gain of vertices induced by the move.

\begin{center}
\vspace{1cm}
\(P=p(N_{11}-3N_6)\)
\\
\vspace{0.5cm}
\begin{tikzpicture}
\draw[thick][red]  (8.,0) -- (9,0);
\node[circle,minimum size=0.8cm,draw=black,fill=white] at (0.,0.0) {11};
\node[circle,minimum size=0.8cm,draw=black,fill=white] at (1.,1.0) {11};
\node[circle,minimum size=0.8cm,draw=black,fill=white] at (1.,0.0) {11};
\node[circle,minimum size=0.8cm,draw=black,fill=white] at (1.,-1.0) {11};
\node[circle,minimum size=0.8cm,draw=black,fill=white] at (2.,2.0) {11};
\node[circle,minimum size=0.8cm,draw=black,fill=white] at (2.,1.0) {11};
\node[circle,minimum size=0.8cm,draw=red,fill=white] at (2.,0.0) {11};
\node[circle,minimum size=0.8cm,draw=black,fill=white] at (2.,-1.0) {11};
\node[circle,minimum size=0.8cm,draw=black,fill=white] at (2.,-2.0) {11};
\node[circle,minimum size=0.8cm,draw=black,fill=white] at (3.,1.0) {11};
\node[circle,minimum size=0.8cm,draw=black,fill=lightgray] at (3.,0.0) {11};
\node[circle,minimum size=0.8cm,draw=black,fill=white] at (3.,-1.0) {11};
\node[circle,minimum size=0.8cm,draw=black,fill=white] at (4.,0.0) {11};
\draw [very thick] [ ->] (4.6,0.0) -- (5.4,0.0);
\node[circle,minimum size=0.8cm,draw=black,fill=white] at (6.,0.0) {11};
\node[circle,minimum size=0.8cm,draw=black,fill=white] at (7.,1.0) {11};
\node[circle,minimum size=0.8cm,draw=black,fill=white] at (7.,0.0) {11};
\node[circle,minimum size=0.8cm,draw=black,fill=white] at (7.,-1.0) {11};
\node[circle,minimum size=0.8cm,draw=black,fill=white] at (8.,2.0) {11};
\node[circle,minimum size=0.8cm,draw=black,fill=white] at (8.,1.0) {11};
\node[circle,minimum size=0.8cm,draw=red,fill=lightgray] at (8.,0.0) {6};
\node[circle,minimum size=0.8cm,draw=black,fill=white] at (8.,-1.0) {11};
\node[circle,minimum size=0.8cm,draw=black,fill=white] at (8.,-2.0) {11};
\node[circle,minimum size=0.8cm,draw=black,fill=white] at (9.,1.0) {11};
\node[circle,minimum size=0.8cm,draw=black,fill=lightgray] at (9.,0.0) {6};
\node[circle,minimum size=0.8cm,draw=black,fill=white] at (9.,-1.0) {11};
\node[circle,minimum size=0.8cm,draw=black,fill=white] at (10.,0.0) {11};
\end{tikzpicture}
\\
\vspace{.5cm}
\(-2N_{11},+2N_6\)
\vspace{0.25cm}
\end{center}

In a similar way, the probability of all the other 15 possible moves (to order $p^2$) are
computed in the Appendix.

\subsection{The master equations}

Collecting all the contributions for each of the probabilities $N_a$
we can now build the master equations governing their evolution in 
this approximation
\begin{align}
\dot{N}_{11}&=-N_{11}\frac{12}{4e^{\beta}+q-4}-2N_{11}p+2N_6-\frac{7}{4}N_6p-2N_{3a}+2N_{10b}+2N_{10a} \\
&+2[(N_{3b}+N_{3a})P_{3\to 10}-(N_{10b}+N_{10a})P_{10\to 3}] \\
\dot{N}_{6}&=2N_{11}p-2N_6+\frac{1}{2}N_6p+4(N_{3a}+N_{3b})+2(N_{3c}P_{3\to 10}-N_{10c}P_{10\to 3}) \\
&-2[(N_{3b}+N_{3a})P_{3\to 10}-(N_{10b}+N_{10a})P_{10\to 3}] \\
\dot{N}_{3a}&=\frac{1}{4}N_6p-2N_{3a}-(N_{3a}P_{3\to 10}-N_{10a}P_{10\to 3}) \\
\dot{N}_{3b}&=\frac{1}{2}N_6p-2N_{3b}-(N_{3b}P_{3\to 10}-N_{10b}P_{10\to 3})+(N_{3c}P_{3\to 10}-N_{10c}P_{10\to 3}) \\
\dot{N}_{3c}&=-4(N_{3c}P_{3\to 10}-N_{10c}P_{10\to 3}) \\
\dot{N}_{10a}&=N_{11}\frac{4}{4e^{\beta}+q-4}-2N_{10a}+(N_{3a}P_{3\to 10}-N_{10a}P_{10\to 3}) \\
\dot{N}_{10b}&=N_{11}\frac{8}{4e^{\beta}+q-4}-2N_{10b}+2(N_{3b}P_{3\to 10}-N_{10b}P_{10\to 3}) \\
\dot{N}_{10c}&=\dot N_{3b}=\frac{1}{2}N_6p-2N_{3b}-(N_{3b}P_{3\to 10}-N_{10b}P_{10\to 3})+(N_{3c}P_{3\to 10}-N_{10c}P_{10\to 3}) .
\end{align}
We want to solve the equations at stationarity, to do so we write down the probabilities in powers of \(p\)
\begin{equation}
\label{eq:power_p}
\begin{aligned}
N_{11}&=\alpha_0+\alpha_1p+\alpha_2p^2 \\
N_{6}&=\beta_1p+\beta_2p^2 \\
N_{3a}&=\gamma_{2a}p^2 \\
N_{3b}&=\gamma_{2b}p^2 \\
N_{3c}&=\gamma_{2c}p^2 \\
N_{10a}&=\delta_{2a}p^2 \\
N_{10b}&=\delta_{2b}p^2 \\
N_{10c}&=\delta_{2c}p^2 . 
\end{aligned}
\end{equation}
The normalization condition \(N_{11}+N_{6}+N_{3a}+N_{3b}+N_{3c}+N_{10a}+N_{10c}+N_{10b}=1\)
implies \(\alpha_0=1\), 
\(\beta_1=-\alpha_1\), \(\alpha_2=-(\beta_2+\gamma_{2a}+\gamma_{2b}+\gamma_{2c}+\delta_{2a}+\delta_{2b}+\delta_{2c})\).
Plugging the expressions in~(\ref{eq:power_p}) in the master equation 
we find from \(\dot{N}_{3c}=0\) that \(\gamma_{2c}=\delta_{2c}P_{10\to3}/P_{3\to10}\),
the first two equations contain first power terms of the form 
\(2\alpha_1p + 2p\), thus \(\alpha_1=-1\) and by construction \(\delta_{2c}=\gamma_{2b}\).
We are left with
\begin{equation}
\begin{aligned}
\dot N_{11}&=-12xp^2+2p^2+2\beta_2p^2 -\frac{7}{4}p^2-2\gamma_{2a}p^2+2\delta_{2b}p^2+\delta_{2a}p^2+\\
&+2p^2[(\gamma_{2b}+\gamma_{2a})P_{3\to 10}-(\delta_{2b}+\delta_{2a})P_{10\to 3}] \\
\dot N_{6}&=-2p^2-2\beta_2p^2+\frac{1}{2}p^2+4p^2(\gamma_{2a}+\gamma_{2b}) \\
&-2p^2[(\gamma_{2b}+\gamma_{2a})P_{3\to 10}-(\delta_{2b}+\delta_{2a})P_{10\to 3}] \\
\dot N_{3a}&=\frac{1}{4}p^2-2\gamma_{2a}p^2-p^2(\gamma_{2a}P_{3\to 10}-\delta_{2a}P_{10\to 3}) \\
\dot N_{3b}&=\frac{1}{2}p^2-2\gamma_{2b}p^2-p^2(\gamma_{2b}P_{3\to 10}-\delta_{2b}P_{10\to 3}) \\
\dot N_{10a}&=4xp^2-2\delta_{2a}p^2+p^2(\gamma_{2a}P_{3\to 10}-\delta_{2a}P_{10\to 3}) \\
\dot N_{10b}&=8xp^2-2\delta_{2b}p^2+2p^2(\gamma_{2a}P_{3\to 10}-\delta_{2a}P_{10\to 3}) ,
\end{aligned}
\end{equation}
where \(x\equiv p^{-2}/(4e^{\beta}+q-4)\). \\
From \(\dot N_{10a}=0\) we get
\begin{equation}
\delta_{2a}=\frac{4x+\gamma_{2a}P_{3\to10}}{2+P_{10\to3}},
\end{equation}
from \(\dot N_{3a}=0\)
\begin{equation}
\gamma_{2a}=\frac{1/2+P_{10\to3}/4+4xP_{10\to3}}{4+2P_{10\to3}+2P_{3\to10}},
\end{equation}
\(\dot N_{10b}=0\) gives
\begin{equation}
\delta_{2b}=\frac{4x+\gamma_{2b}P_{3\to10}}{1+P_{10\to3}},
\end{equation}
\(\dot N_{10b}=0\)
\begin{equation}
\gamma_{2b}=\frac{1/2+P_{10\to3}/2+4xP_{10\to3}}{2+2P_{10\to3}+P_{3\to10}},
\end{equation}
and finally from \(\dot N_6=0\)
\begin{equation}
\beta_2=-3/4+2(\gamma{21}+\gamma{22})-[(\gamma_{2a}+\gamma_{2b})P_{3\to 10}-(\delta_{2b}+\delta_{2a})P_{10\to 3}] .
\end{equation}
Thus summarizing
\begin{equation}
\begin{aligned}
\alpha_0&=1 \\
\alpha_1&=-1 \\
\alpha_2&=-(\beta_2+\gamma_{2a}+\gamma_{2b}+\gamma_{2c}+\delta_{2a}+\delta_{2b}+\delta_{2c}) \\
\beta_1&=-\alpha_1 \\
\beta_2&=-3/4+2(\gamma{21}+\gamma{22})-[(\gamma_{2a}+\gamma_{2b})P_{3\to 10}-(\delta_{2b}+\delta_{2a})P_{10\to 3}] \\
\gamma_{2a}&=\frac{1/2+P_{10\to3}/4+4xP_{10\to3}}{4+2P_{10\to3}+2P_{3\to10}} \\
\gamma_{2b}&=\frac{1/2+P_{10\to3}/2+4xP_{10\to3}}{2+2P_{10\to3}+P_{3\to10}} \\
\gamma_{2c}&=\delta_{2c}P_{10\to3}/P_{3\to10} \\
\delta_{2a}&=\frac{4x+\gamma_{2a}P_{3\to10}}{2+P_{10\to3}} \\
\delta_{2b}&=\frac{4x+\gamma_{2b}P_{3\to10}}{1+P_{10\to3}}\\
\delta_{2c}&=\gamma_{2b}.
\end{aligned}
\end{equation}

\subsection{Numerical tests}

In order to put the approach above to the numerical test, 
we collected the proportions $N_{a}$ measured with the heat bath Monte Carlo simulations 
and we compared them to the values computed with the 
master equation analysis. Concretely, 
we used systems with $L=10^3$, and  $q=10^4, \ 10^5$ and $10^6$, at
$T/T_c=0.99$.  The numerical and analytic data are displayed in Tab.~\ref{sum10k100k1m}. 
The number of digits shown correspond to results up to order $p^2$.  The agreement between the 
values found with the two approaches is excellent.

\begin{table}[h!]
\centering
\begin{tabular}{ |c  ||  c | c || c | c|| c | c |} 
\hline
   $q $    &   \multicolumn{2}{ c|| }{10 000}  & \multicolumn{2}{c||}{100 000}  & \multicolumn{2}{c|}{1 000 000}   \\
\hline
             & numerical & analytic & numerical & analytic & numerical & analytic  \\
\hline             
$N_{11}$ & 0.95731 &  0.95729 &   0.986509  &  0.986509 & 0.9957020   &  0.9957023 \\
$N_6$ &     0.04054 &   0.04064 &  0.013269  &  0.013272 &  0.0042752   &  0.0042751 \\
$N_{3a}$ & 0.00021 &  0.00021 &   0.000022  &  0.000022 &  0.0000023   &  0.0000023 \\
$N_{3b}$ & 0.00042 &   0.00041 &  0.000044  &  0.000044 &  0.0000046   &  0.0000046 \\
$N_{3c}$ & 0.00048 &    0.00046 & 0.000050  &  0.000050 &  0.0000053   &  0.0000053 \\
$N_{10a}$ & 0.00019 &  0.00019 & 0.000020  &  0.000020 &  0.0000020   &  0.0000020 \\
$N_{10b}$ & 0.00044 &  0.00041 & 0.000045  &  0.000044 &  0.0000046   &  0.0000046 \\
$N_{10c}$ & 0.00037 &  0.00038 & 0.000039  &  0.000039 &  0.0000040   &  0.0000040 \\
\hline
\end{tabular}
\caption{$N_{a}$ for systems with $ L=10^3$ and $q= 10^4, \ 10^5, \ 10^6$ 
evolving at temperatures $T/T_c=0.99$ after an instantaneous quench from infinite temperature.
The first column show the numerical values  at MC times such that the system is stationary
in the metastable state, 
while the second ones give the asymptotic values calculated with the master equations approach. 
Only the relevant values (up to order $p^2$) are shown. 
The  error bars on the numerical values are always smaller than one on the last shown digits.}
\label{sum10k100k1m}
\end{table}

In Tab.~\ref{sum3} we show data for a system with linear size $L=10^3$
and $q=10^6$, and we vary the temperature, moving progressively towards criticality at $T_c$. 
As explained below, for this value of $q$, we observe a divergency 
of the time required to reach a ferromagnetic state at $T/T_c \simeq 0.9$. 
The data in Tab.~\ref{sum3} show that the analytic approximation is very good (in the metastable state)
even moderately away from $T_c$. However, the numerical measurements at $T/T_c =0.88$ 
have been done at time $t=10^3$, and at this time the agreement between numerical
and analytical data is still good but not as good as for the higher temperatures. In particular, one can notice 
a relatively important difference in $N_{11}$ and $N_{3c}$. For longer measuring times, one would see
this difference increase, showing that the system leaves the metastable state at $T/T_c =0.88$.
For the higher temperatures, there are no time-dependencies in the numerical results and for all 
purposes the metastable states remain for ever.

\begin{table}[h!]
\centering
\begin{tabular}{ |c |c|c|| @{}c@{} | @{}c@{} | @{}c@{}| @{}c@{} | @{}c@{} | @{}c@{} | @{}c@{} | } 
\hline
   $T/T_c$ & $p$  &   &$\quad \quad N_{11}\quad \quad$ & $\quad \quad N_6\quad \quad $ & $\,10^3 N_{3a}\,$ & $\,10^3 N_{3b}\,$  &   $\,10^3 N_{3c}\,$ &  $\,10^3 N_{10a}\,$ &  $\,10^3 N_{10c}\,$ \\
\hline
\multirow{2}{*}{0.88} &\multirow{2}{*}{0.01017}    &numeric& 0.9895816 & 0.0101646& 0.0130 &0.0260 & 0.1772&0.0020& 0.0039 \\
   &&  analytic&         0.9895916 &  0.0101674 &  0.0129 & 0.0259 & 0.1705 & 0.0020  &  0.0039   \\
  \hline
\multirow{2}{*}{0.92}&\multirow{2}{*}{0.00725} & numeric &  0.9926679  & 0.0072481 & 0.0066 &  0.0132 & 0.0444 & 0.0020 & 0.0039  \\
  && analytic   &0.9926690 &  0.0072485 &  0.0066 & 0.0131  & 0.0438  & 0.0020  &  0.0040   \\
 \hline
\multirow{2}{*}{0.98}&\multirow{2}{*}{0.00459} &numeric &0.9953845 &  0.0045892 &   0.0026 & 0.0053 & 0.0070 & 0.0020 & 0.0040   \\
    && analytic  &0.9953847  &  0.0045892 &  0.0026 & 0.0053  & 0.0070  & 0.0020  &  0.0040  \\ 
 \hline
\multirow{2}{*}{0.99}&\multirow{2}{*}{0.00428} &numeric &0.9957020 & 0.0042752 & 0.0023  & 0.0046 & 0.0053 & 0.0020& 0.0040  \\
    && analytic &0.9957023 & 0.0042751 & 0.0023  & 0.0046 & 0.0053 &  0.0020 &  0.0040  \\
\hline
\end{tabular}
\caption{$N_{a}$ for systems with linear size $L=10^3$, $q=10^6$ and various values of $T/T_c$ (corresponding
to different values of $p$ ((second row)). 
For each temperature, the first line shows 
the numerical values  at MC times such that the system is stationary
in the metastable state, 
while the second ones give the asymptotic values calculated with the master equations approach. 
The error bars for the numerical values are of the order the last digit or smaller and 
they are not shown. We  also have MC data for $N_1$, $10^3 N_1 = 0.0044$ at $T/T_c=0.88$, 
$10^3 N_1 = 0.0005$ at $T/T_c=0.92$, $10^3 N_1 = 0.0000$ at $T/T_c=0.98$ and $T/T_c=0.99$.
}
\label{sum3}
\end{table}

\begin{figure}[t!]
\begin{center}
\scalebox{.7}{\input{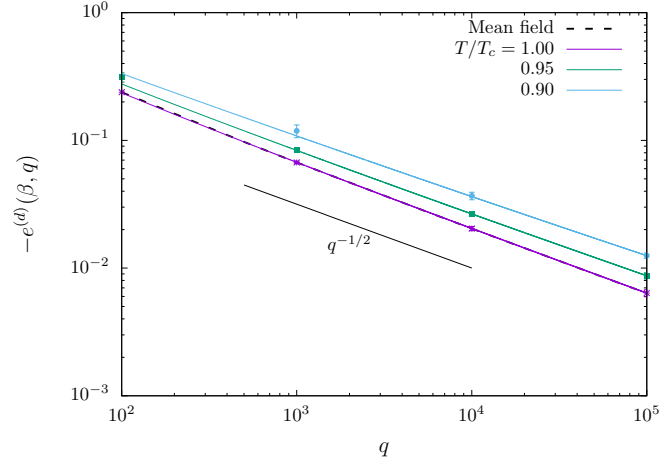}}
\hspace{1cm}
\end{center}
\caption{Theoretical predictions in Eq.~(\ref{eq:plateau}) and simulations results for the
energy density of a system with \(L=200\) when it is stuck in a paramagnetic metastable configuration,
as a function of the number of states \(q\),  
for several ratios of the quench temperature over the critical one.
The numerical values are time averages over  a single run.
The error bars equal a standard deviation. 
The dashed tilted line correspond to the mean field exact result at criticality~\cite{Wu82}.}
\label{fig:EnPlatDMF}
\end{figure}

\begin{figure}[h!]
\begin{center}
\scalebox{.7}{\input{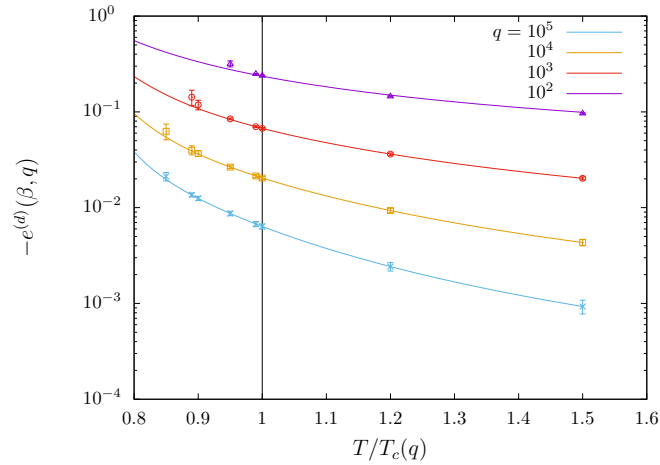}}
\hspace{1cm}
\end{center}
\caption{Energy density of the disordered  metastable state
\textit{vs} \(T/T_c\) for several values of \(q\) (increasing from bottom to top), 
evaluated from Eq.~(\ref{eq:plateau}) (colored solid lines). 
Values from simulations are also presented with data points. They are time averages
of the energy density. The error bars correspond to a standard deviation.
The critical temperature is indicated with a vertical black line.}
\label{fig:EvsTD}
\end{figure}

Once the proportions \(N_a\) are known it is possible to 
thermodynamically characterize  the metastable states. For instance, we can evaluate 
the energy per spin of the disordered metastable state extended below the critical temperature,
exploiting the stationary solutions obtained above.
The only configurations that contribute to the energy are the \((6)\) ones with one bond and the 
\((3)\) ones with two bonds. Thus we have
\begin{equation}
\label{eq:plateau}
e^{(d)}(\beta,q)=-\frac{1}{2}(N_6(\beta,q)+2N_3(\beta,q))
\; ,
\end{equation}
where the \(1/2\) factor avoids double counting of the bonds on the lattice.
Note that for quench inverse temperature \(\beta < \beta_c\) the expression 
in Eq.~(\ref{eq:plateau}) should provide the equilibrium value of the energy at \(\beta\).
In Fig.~\ref{fig:EnPlatDMF} we plot the energy density of the disordered state
as predicted by Eq.~(\ref{eq:plateau}) as a function of \(q\) at different ratios
between the quench temperature and the critical one. The values of the energy density obtained with  Monte Carlo 
simulations are also reported in the figure.
The latter are time averages over single runs computed as long as the system
stays in the metastable state  (the error bars represent one standard deviation).
A comparison with the exact mean field result for the energy at criticality~\cite{Wu82} is reported.
It is possible to appreciate that, for all temperatures,
the energy decreases (in absolute value) approximatively as \(q^{-1/2}\),
this is expected because the major contribution to Eq.~(\ref{eq:plateau})
is given by the \((6)\) term which scales indeed as \(q^{-1/2}\) (see section above).
Figure~\ref{fig:EvsTD} shows instead the behaviour of the energy density of the disordered state 
as a function of  the final quench temperature.
The results of the expansion are again tested against Monte Carlo numerical simulations 
showing really good agreement.

\section{Upper-critical quenches: the ordered metastable phase}
\label{sec:supercritical}

As we anticipated above, the upper-critical protocol, which deals with the persistence of the ordered
phase after a quench to a temperature \(T>T_c\) starting from a fully ordered configuration,
is less interesting from a technical point of view. We nonetheless perform 
a similar analysis (though less rich in terms of numerical evaluations) as for the disordered phase
in order to complete the picture of metastability.

\subsection{Large \(q\) and large \(N\) behaviour}

Let us take the initial configuration to be at 
zero temperature, that is to say, a completely ordered state.
Thus, the system is in one of the \(q\) possible ground states and, consequently,   
all the \(N\) sites are in state (0).

Recalling that (see Eq.~(\ref{eq:beta_q_relation})) for large \(q\) we have \(e^{\beta}\simeq q^{T_c/2T}\),
during a lattice update, the probability for a state (0) to turn into a state (7) can be written as
\begin{equation}
\label{R116_ordered}
P_{0 \rightarrow 7} = \frac{q-1}{q+e^{4\beta}-1} \simeq \frac{q}{q + q^{2T_c/T}}
=\frac{1}{1 + q^{2T_c/T-1}} \ .
\end{equation}
Thus, in the upper critical regime,  the crossover temperature that separates two very different behaviours 
in the \(q\rightarrow \infty\) limit
is \(T=2T_c\):
\begin{equation}
P_{0 \rightarrow 7} \rightarrow 1 
\qquad\quad
\mbox{at}
\qquad\quad
T>2T_c 
\; , 
\end{equation}
the (0) states turn into (7) states, and the system disorders really fast.
At the crossover temperature
\begin{equation}
P_{0 \rightarrow 7} \rightarrow 1/2
\qquad\quad
\mbox{at}
\qquad\quad
T=2T_c 
\; , 
\end{equation}
implying that states (7) can appear. Every (7) states will have as neighbours (1) states which
(always in the limit \(q\rightarrow\infty\)) will  become states (8) with probability \(P_{1\rightarrow 8}\rightarrow 1\),
and bring the system to a disordered configuration.
Finally, 
\begin{equation}
P_{0 \rightarrow 7} \rightarrow 0 
\qquad\quad
\mbox{at}
\qquad\quad
T<2T_c 
\; , 
\end{equation}
and the state (0) is completely stable in this temperature window close to $T_c$.

\begin{figure}[h!]
\begin{center}
\scalebox{.7}{\input{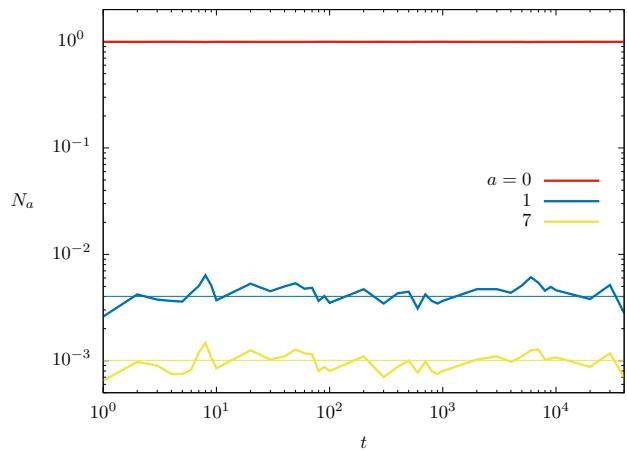}
\hspace{1cm}
}
\end{center}
\caption{$N_a(t)$ for $a=0, 1, 7$ evolving in time at $T=1.01 \, T_c$ in a square lattice system with linear size 
$L=10^3$ and $q=10^3$.
In thin lines are reported the analytical predictions obtained from the master equations below,
in thick lines data from a numerical simulation. Note that the (1) and (7) abundances are one the vertical translation of the other.
This is due to the fact that, by construction, there are  four (1) states for every (7) one (see below).}
\label{fig:UpperWeights}
\end{figure}

Going back to large but finite \(q\), in Fig.~\ref{fig:UpperWeights}, we show the evolution of $N_a$ as a function 
of time for $a=0, 1$ and $7$, we only show the states which contribute the most.

Therefore, at upper critical  temperatures, the following hierarchy holds
\begin{equation}
N_1\simeq N_7 \ll N_0\simeq 1 \ ,
\end{equation}
where the $N_a$ are normalised by the number of spins in the sample, and all other states are negligible.

\subsection{The leading updates at \(T\simeq T_c\)}

Using again the expansion parameter \(p\) with, 
\begin{equation}
p^2\simeq q^{-1}\simeq P_{0\rightarrow 7}
\qquad\quad
\mbox{at}
\qquad\quad
T \sim T_c 
\; ,
\end{equation}
we consider the evolution of $N_{0} \simeq 1 , N_{7} \simeq p^2$ and $N_{1} \simeq p^2$.
Again we stop at second order in \(p\).

It is straightforward to verify that the only structure that can appear in the sea of
aligned spins ({\it i.e.}, in the (0) state), with a probability proportional to 
\(p^2\) or greater, is a (7) state surrounded by (1) states
\begin{center}
\begin{tikzpicture}
\draw[thick][red]  (0,0) -- (1,0);
\draw[thick][red]  (3,0) -- (4,0);
\draw[thick][red]  (2,1.) -- (2.,2.);
\draw[thick][red]  (2,-1.) -- (2.,-2.);
\draw[thick][red]  (1,0.) -- (1.,1.);
\draw[thick][red]  (1,0.) -- (1.,-1.);
\draw[thick][red]  (3,0.) -- (3.,1.);
\draw[thick][red]  (3,0.) -- (3.,-1.);
\draw[thick][red]  (1,1.) -- (2.,1.);
\draw[thick][red]  (2,1.) -- (3,1.);
\draw[thick][red]  (1,-1.) -- (2,-1.);
\draw[thick][red]  (2,-1.) -- (3.,-1.);
\node[circle,minimum size=0.8cm,draw=black,fill=lightgray] at (0.,0.0) {0};
\node[circle,minimum size=0.8cm,draw=black,fill=lightgray] at (1.,1.0) {0};
\node[circle,minimum size=0.8cm,draw=black,fill=lightgray] at (1.,0.0) {1};
\node[circle,minimum size=0.8cm,draw=black,fill=lightgray] at (1.,-1.0) {0};
\node[circle,minimum size=0.8cm,draw=black,fill=lightgray] at (2.,2.0) {0};
\node[circle,minimum size=0.8cm,draw=black,fill=lightgray] at (2.,1.0) {1};
\node[circle,minimum size=0.8cm,draw=black,fill=white] at (2.,0.0) {7};
\node[circle,minimum size=0.8cm,draw=black,fill=lightgray] at (2.,-1.0) {1};
\node[circle,minimum size=0.8cm,draw=black,fill=lightgray] at (2.,-2.0) {0};
\node[circle,minimum size=0.8cm,draw=black,fill=lightgray] at (3.,1.0) {0};
\node[circle,minimum size=0.8cm,draw=black,fill=lightgray] at (3.,0.0) {1};
\node[circle,minimum size=0.8cm,draw=black,fill=lightgray] at (3.,-1.0) {0};
\node[circle,minimum size=0.8cm,draw=black,fill=lightgray] at (4.,0.0) {0};
\end{tikzpicture}
\vspace{0.5cm}
\end{center}
Indeed there are only two ways to build different structures from the one above. The first one is that a (1),
which has a probability proportional to \(p^2\) to be picked, turns into a (4) or into an (8),
respectively with probabilities \(P_{1\rightarrow4}\sim p^2\) and \(P_{1\rightarrow4}\sim p\).
The other possibility is that a (0) close to a 1, which again has probability proportional to \(p^2\)
to be picked, turns into a (7), with probability \(P_{0\rightarrow7}\sim p^2\).
The overall probabilities therefore are such that both scenarios are negligible in our approximation.

The only moves that should  be taken into account to build a master equation for the ordered case
are the switching of a (0) (surrounded by other (0) states) into a (7) and \textit{vice versa}.
In particular, we have that the probability of picking such a (0) is \(N_0-8N_7\), because there are 8 (0)
states next to a (1) surrounding each (7), but to the second order in \(p\) we only retain \(N_0\), and the probability 
for it to turn into a (7) creating in doing so also 4 (1) states is \(P_{0\rightarrow7}\)

\begin{center}
\nopagebreak
\(P=N_{0}P_{0\rightarrow7}\)
\vspace{0.25cm}
\nopagebreak
\\
\nopagebreak
\begin{tikzpicture}
\draw[thick][red]  (0,0) -- (1,0);
\draw[thick][red]  (3,0) -- (4,0);
\draw[thick][red]  (2,1.) -- (2.,2.);
\draw[thick][red]  (2,-1.) -- (2.,-2.);
\draw[thick][red]  (1,0.) -- (1.,1.);
\draw[thick][red]  (1,0.) -- (1.,-1.);
\draw[thick][red]  (2,0.) -- (2.,1.);
\draw[thick][red]  (2,0.) -- (2.,-1.);
\draw[thick][red]  (1,0.) -- (2.,0.);
\draw[thick][red]  (2,0.) -- (3.,0.);
\draw[thick][red]  (3,0.) -- (3.,1.);
\draw[thick][red]  (3,0.) -- (3.,-1.);
\draw[thick][red]  (1,1.) -- (2.,1.);
\draw[thick][red]  (2,1.) -- (3,1.);
\draw[thick][red]  (1,-1.) -- (2,-1.);
\draw[thick][red]  (2,-1.) -- (3.,-1.);
\draw[thick][red]  (9,0) -- (10,0);
\draw[thick][red]  (8,1.) -- (8.,2.);
\draw[thick][red]  (8,-1.) -- (8.,-2.);
\draw[thick][red]  (9,0.) -- (9.,1.);
\draw[thick][red]  (9,0.) -- (9.,-1.);
\draw[thick][red]  (7,1.) -- (8.,1.);
\draw[thick][red]  (8,1.) -- (9,1.);
\draw[thick][red]  (7,-1.) -- (8,-1.);
\draw[thick][red]  (8,-1.) -- (9.,-1.);
\draw[thick][red]  (6,0.) -- (7.,0.);
\draw[thick][red]  (7,0.) -- (7.,1.);
\draw[thick][red]  (7,0.) -- (7.,-1.);
\node[circle,minimum size=0.8cm,draw=black,fill=lightgray] at (0.,0.0) {0};
\node[circle,minimum size=0.8cm,draw=black,fill=lightgray] at (1.,1.0) {0};
\node[circle,minimum size=0.8cm,draw=black,fill=lightgray] at (1.,0.0) {0};
\node[circle,minimum size=0.8cm,draw=black,fill=lightgray] at (1.,-1.0) {0};
\node[circle,minimum size=0.8cm,draw=black,fill=lightgray] at (2.,2.0) {0};
\node[circle,minimum size=0.8cm,draw=black,fill=lightgray] at (2.,1.0) {0};
\node[circle,minimum size=0.8cm,draw=red,fill=lightgray] at (2.,0.0) {0};
\node[circle,minimum size=0.8cm,draw=black,fill=lightgray] at (2.,-1.0) {0};
\node[circle,minimum size=0.8cm,draw=black,fill=lightgray] at (2.,-2.0) {0};
\node[circle,minimum size=0.8cm,draw=black,fill=lightgray] at (3.,1.0) {0};
\node[circle,minimum size=0.8cm,draw=black,fill=lightgray] at (3.,0.0) {0};
\node[circle,minimum size=0.8cm,draw=black,fill=lightgray] at (3.,-1.0) {0};
\node[circle,minimum size=0.8cm,draw=black,fill=lightgray] at (4.,0.0) {0};
\draw [very thick] [ ->] (4.6,0.0) -- (5.4,0.0);
\node[circle,minimum size=0.8cm,draw=black,fill=lightgray] at (6.,0.0) {0};
\node[circle,minimum size=0.8cm,draw=black,fill=lightgray] at (7.,1.0) {0};
\node[circle,minimum size=0.8cm,draw=black,fill=lightgray] at (7.,0.0) {1};
\node[circle,minimum size=0.8cm,draw=black,fill=lightgray] at (7.,-1.0) {0};
\node[circle,minimum size=0.8cm,draw=black,fill=lightgray] at (8.,2.0) {0};
\node[circle,minimum size=0.8cm,draw=black,fill=lightgray] at (8.,1.0) {1};
\node[circle,minimum size=0.8cm,draw=red,fill=white] at (8.,0.0) {7};
\node[circle,minimum size=0.8cm,draw=black,fill=lightgray] at (8.,-1.0) {1};
\node[circle,minimum size=0.8cm,draw=black,fill=lightgray] at (8.,-2.0) {0};
\node[circle,minimum size=0.8cm,draw=black,fill=lightgray] at (9.,1.0) {0};
\node[circle,minimum size=0.8cm,draw=black,fill=lightgray] at (9.,0.0) {1};
\node[circle,minimum size=0.8cm,draw=black,fill=lightgray] at (9.,-1.0) {0};
\node[circle,minimum size=0.8cm,draw=black,fill=lightgray] at (10.,0.0) {0};
\end{tikzpicture}
\\
\(-5N_{0},+N_{7},+4N_{1}\)
\vspace{0.5cm}
\end{center}
The inverse move, consistently, with probability \(N_7P_{7\rightarrow0}\) causes the destruction
of 4 (1) states and of 1 (7) state creating 5 (0) states 
\begin{center}
\(P=N_{7}P_{7\rightarrow0}\)
\vspace{0.25cm}
\\
\begin{tikzpicture}
\draw[thick][red]  (0,0) -- (1,0);
\draw[thick][red]  (3,0) -- (4,0);
\draw[thick][red]  (2,1.) -- (2.,2.);
\draw[thick][red]  (2,-1.) -- (2.,-2.);
\draw[thick][red]  (1,0.) -- (1.,1.);
\draw[thick][red]  (1,0.) -- (1.,-1.);
\draw[thick][red]  (3,0.) -- (3.,1.);
\draw[thick][red]  (3,0.) -- (3.,-1.);
\draw[thick][red]  (1,1.) -- (2.,1.);
\draw[thick][red]  (2,1.) -- (3,1.);
\draw[thick][red]  (1,-1.) -- (2,-1.);
\draw[thick][red]  (2,-1.) -- (3.,-1.);
\draw[thick][red]  (9,0) -- (10,0);
\draw[thick][red]  (8,1.) -- (8.,2.);
\draw[thick][red]  (8,-1.) -- (8.,-2.);
\draw[thick][red]  (9,0.) -- (9.,1.);
\draw[thick][red]  (9,0.) -- (9.,-1.);
\draw[thick][red]  (7,1.) -- (8.,1.);
\draw[thick][red]  (8,1.) -- (9,1.);
\draw[thick][red]  (7,-1.) -- (8,-1.);
\draw[thick][red]  (8,-1.) -- (9.,-1.);
\draw[thick][red]  (6,0.) -- (7.,0.);
\draw[thick][red]  (7,0.) -- (7.,1.);
\draw[thick][red]  (7,0.) -- (7.,-1.);
\draw[thick][red]  (8,0.) -- (9.,0.);
\draw[thick][red]  (8,0.) -- (8.,-1.);
\draw[thick][red]  (8,0.) -- (8.,1.);
\draw[thick][red]  (7,0.) -- (8.,0.);
\node[circle,minimum size=0.8cm,draw=black,fill=lightgray] at (0.,0.0) {0};
\node[circle,minimum size=0.8cm,draw=black,fill=lightgray] at (1.,1.0) {0};
\node[circle,minimum size=0.8cm,draw=black,fill=lightgray] at (1.,0.0) {1};
\node[circle,minimum size=0.8cm,draw=black,fill=lightgray] at (1.,-1.0) {0};
\node[circle,minimum size=0.8cm,draw=black,fill=lightgray] at (2.,2.0) {0};
\node[circle,minimum size=0.8cm,draw=black,fill=lightgray] at (2.,1.0) {1};
\node[circle,minimum size=0.8cm,draw=red,fill=white] at (2.,0.0) {7};
\node[circle,minimum size=0.8cm,draw=black,fill=lightgray] at (2.,-1.0) {1};
\node[circle,minimum size=0.8cm,draw=black,fill=lightgray] at (2.,-2.0) {0};
\node[circle,minimum size=0.8cm,draw=black,fill=lightgray] at (3.,1.0) {0};
\node[circle,minimum size=0.8cm,draw=black,fill=lightgray] at (3.,0.0) {1};
\node[circle,minimum size=0.8cm,draw=black,fill=lightgray] at (3.,-1.0) {0};
\node[circle,minimum size=0.8cm,draw=black,fill=lightgray] at (4.,0.0) {0};
\draw [very thick] [ ->] (4.6,0.0) -- (5.4,0.0);
\node[circle,minimum size=0.8cm,draw=black,fill=lightgray] at (6.,0.0) {0};
\node[circle,minimum size=0.8cm,draw=black,fill=lightgray] at (7.,1.0) {0};
\node[circle,minimum size=0.8cm,draw=black,fill=lightgray] at (7.,0.0) {0};
\node[circle,minimum size=0.8cm,draw=black,fill=lightgray] at (7.,-1.0) {0};
\node[circle,minimum size=0.8cm,draw=black,fill=lightgray] at (8.,2.0) {0};
\node[circle,minimum size=0.8cm,draw=black,fill=lightgray] at (8.,1.0) {0};
\node[circle,minimum size=0.8cm,draw=red,fill=lightgray] at (8.,0.0) {0};
\node[circle,minimum size=0.8cm,draw=black,fill=lightgray] at (8.,-1.0) {0};
\node[circle,minimum size=0.8cm,draw=black,fill=lightgray] at (8.,-2.0) {0};
\node[circle,minimum size=0.8cm,draw=black,fill=lightgray] at (9.,1.0) {0};
\node[circle,minimum size=0.8cm,draw=black,fill=lightgray] at (9.,0.0) {0};
\node[circle,minimum size=0.8cm,draw=black,fill=lightgray] at (9.,-1.0) {0};
\node[circle,minimum size=0.8cm,draw=black,fill=lightgray] at (10.,0.0) {0};
\end{tikzpicture}
\\
\(-N_{7},-4N_{1},+5N_{0},\)
\vspace{0.5cm}
\end{center}

\subsection{The master equations}

The master equations are therefore 
\begin{equation}
\begin{aligned}
\dot N_{0}&=-5N_{0}P_{0\rightarrow7} + 5N_{7}P_{7\rightarrow0}\ \\
\dot N_{7}&=-N_{7}P_{7\rightarrow0} + N_{0}P_{0\rightarrow7}\ \ \\
\dot N_{1}&=-4N_{7}P_{7\rightarrow0} + 4N_{0}P_{0\rightarrow7}\ \ .
\end{aligned}
\end{equation}
To solve them we write down the probabilities in powers of \(p\)
\begin{equation}
\label{eq:power_p2}
\begin{aligned}
N_{0}&=\alpha_0+\alpha_1p+\alpha_2p^2 \\
N_{7}&=\beta_2p^2 \\
N_{1}&=\gamma_{2}p^2 . 
\end{aligned}
\end{equation}
By construction we have \(N_1=4N_7\) and so \(\gamma_2=4\beta_2\), moreover the normalization condition \(N_{0}+N_{7}+N_{4}=1\)
impose \(\alpha_0=1\), \(\alpha_1=0\) and \(\alpha_2=-5\beta_2\). Finally, looking for the stationary solution of either one 
of the three differential equations above, we find \(\beta_2=1/P_{7\rightarrow0}\). Summarizing
\begin{equation}
\begin{aligned}
\alpha_0&=1 \ \\
\alpha_1&=0 \ \\
\alpha_2&=-5/P_{7\rightarrow0} \  \\
\beta_2&=1/P_{7\rightarrow0} \ \\
\gamma_{2}&=4/P_{7\rightarrow0} \ .
\end{aligned}
\end{equation}

\subsection{Numerical tests}

\begin{figure}[h!]
\begin{center}
\scalebox{.7}{\input{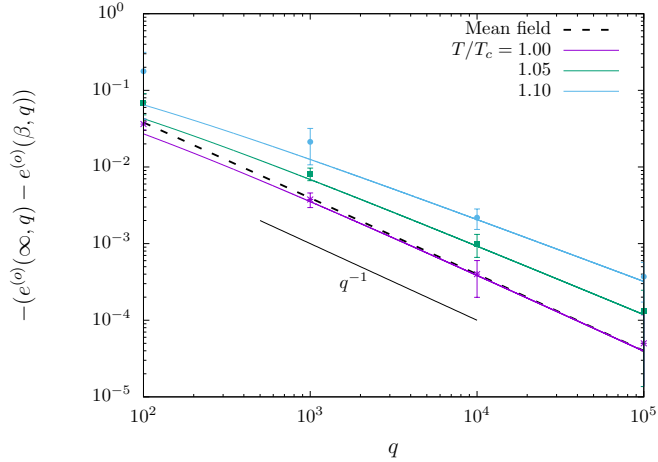}}
\hspace{1cm}
\end{center}
\caption{Energy density of the ordered state 
as predicted by Eq.~(\ref{eq:plateauOrd}) and   simulation data,
for \(L=200\),
as a function of the number of states \(q\), for several ratios of the quench temperature.
The numerical values are averages in time of the energy for a single realisation,
the error bars correspond to a standard deviation. 
Exact mean field predictions at criticality are reported as well (black dashed line)~\cite{Wu82}.}
\label{fig:EnPlatOMF}
\end{figure}

\begin{figure}[h!]
\begin{center}
\scalebox{.7}{\input{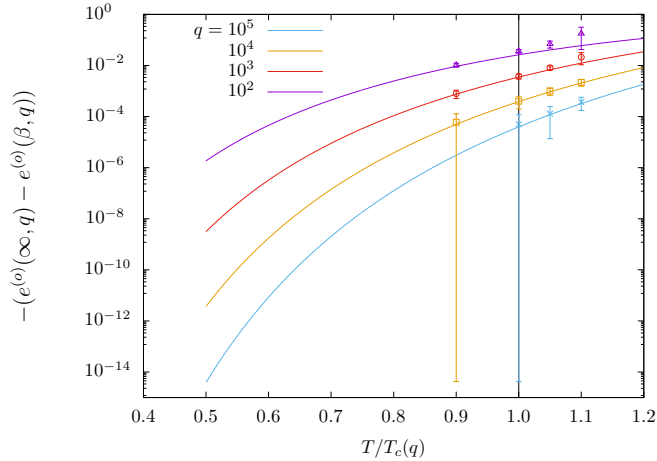}}
\hspace{1cm}
\end{center}
\caption{Ordered energy \textit{vs} \(T/T_c\) for several values of \(q\), evaluated from Eq.~(\ref{eq:plateau}).
Values from simulations for \(L=200\) are also present and are time averages
for a single realization of the energy of the system as long as it stays in the metastable state,
the error bars correspond to a standard deviation.}
\label{fig:EvsTO}
\end{figure}

We can put the results from the previous section to the numerical test analysing, as for the disordered case,
an interesting observable: the energy density of the metastable state.
In this case the spin which falls in the \((0)\) configuration contributes with four bonds,
while the ones in \((4)\) with three bonds. The ordered energy density thus reads
\begin{equation}
\label{eq:plateauOrd}
e^{(o)}(\beta,q)=-\frac{1}{2}(4N_0(\beta,q)+3N_4(\beta,q)) \ .
\end{equation}
This energy scales as \(q^{-1}\) at fixed temperature, consistently with the fact that the major
contribution comes from \((0)\). The agreement with the mean field results~\cite{Wu82} and
the outcome of the simulations
analysed as in the disordered case is really good as can be checked by inspecting Fig.~\ref{fig:EnPlatOMF}.
The dependence of the energy density of the ordered state, 
as evaluated from Eq.~(\ref{eq:plateauOrd}), on temperature is
portrayed in Fig.~\ref{fig:EvsTO}, where the comparison to the results of 
numerical evaluations shows again a perfect agreement.

\section{Conclusions}

Most dynamic studies of the bidimensional Potts model focused on the analysis of the coarsening 
dynamics after deep quenches at moderate subcritical 
temperatures~\cite{Glazier90,Petri08,Loureiro10,Loureiro12} so as to avoid getting stuck
in long-lived metastable configurations~\cite{Lifshitz62,Glazier90,Ferrero07,Ibanez07b,Olejarz13,Denholm}. The 
study of metastability and thermally assisted nucleation close to the critical temperature in this rather simple model 
has not been so much developed in the literature.

Numerical evidence for thermodynamic metastability in finite but large 
size systems with $q>4$ was provided in various papers. In particular, the analysis 
of the short-time dynamics~\cite{Ferrero09} and 
Binder cumulant~\cite{Ferrero11} was recently used with this purpose. However, 
Meunier and Morel~\cite{Meunier00} argued that thermodynamic metastability 
should disappear in the infinite system size limit and other authors~\cite{BerganzaEPL} provided arguments 
supporting this claim.  Extracting the infinite size limit behaviour, 
and the eventual disappearance of metastability from numerical studies 
is, however, a dauntingly hard task.  

Last year, some of us wrote a short note on the nucleation and growth dynamics of the 
two dimensional Potts model~\cite{Corberi19}. With it we started our study of metastability in this (and eventually other)
systems with first order thermal phase transitions. In this paper we developed a large $q$ expansion 
of the heat bath microscopic dynamics 
that allowed us to deduce, analytically, the metastability properties of the finite but large size 
model, in a rather wide range of temperatures around criticality (namely, from $T_c/2$ to $2T_c$). 
 Although in the strictly infinite size limit the spinodals are expected to approach the critical point~\cite{Meunier00}, 
we observe that the lifetime of the metastable state goes beyond reasonable times for relatively small system 
sizes. Our expansion allows us to capture the properties of these metastable states with amazing numerical 
accuracy.

\appendix
\section{Appendix: Probability of the moves}

Consider starting from a state (11) next to a structure \(\textbf A\),  
turn it into a state (6), and make then a structure \(\textbf B\) be born. The probability of picking 
the starting site is \(N_6\) because there are 2 (11) in such position
for every structure \(\textbf A\) (again we are keeping only the terms which at the end will contribute
up to the second order) and the probability to switch to (6) exactly in the needed
direction is \(p/4\). The probability of the move is thus
\(pN_6/4\) and we end up with with 1 (11) less and 1 (3a) more.

\begin{center}
\vspace{1cm}
\(P=pN_{6}/4\)
\\
\begin{tikzpicture}
\draw[thick][red]  (3.,0) -- (4,0);
\draw[thick][red]  (8.,0) -- (9,0);
\draw[thick][red]  (9.,0) -- (10,0);
\node[circle,minimum size=0.8cm,draw=black,fill=white] at (0.,0.0) {11};
\node[circle,minimum size=0.8cm,draw=black,fill=white] at (1.,1.0) {11};
\node[circle,minimum size=0.8cm,draw=black,fill=white] at (1.,0.0) {11};
\node[circle,minimum size=0.8cm,draw=black,fill=white] at (1.,-1.0) {11};
\node[circle,minimum size=0.8cm,draw=black,fill=white] at (2.,2.0) {11};
\node[circle,minimum size=0.8cm,draw=black,fill=white] at (2.,1.0) {11};
\node[circle,minimum size=0.8cm,draw=red,fill=white] at (2.,0.0) {11};
\node[circle,minimum size=0.8cm,draw=black,fill=white] at (2.,-1.0) {11};
\node[circle,minimum size=0.8cm,draw=black,fill=white] at (2.,-2.0) {11};
\node[circle,minimum size=0.8cm,draw=black,fill=white] at (3.,1.0) {11};
\node[circle,minimum size=0.8cm,draw=black,fill=lightgray] at (3.,0.0) {6};
\node[circle,minimum size=0.8cm,draw=black,fill=white] at (3.,-1.0) {11};
\node[circle,minimum size=0.8cm,draw=black,fill=lightgray] at (4.,0.0) {6};
\draw [very thick] [ ->] (4.6,0.0) -- (5.4,0.0);
\node[circle,minimum size=0.8cm,draw=black,fill=white] at (6.,0.0) {11};
\node[circle,minimum size=0.8cm,draw=black,fill=white] at (7.,1.0) {11};
\node[circle,minimum size=0.8cm,draw=black,fill=white] at (7.,0.0) {11};
\node[circle,minimum size=0.8cm,draw=black,fill=white] at (7.,-1.0) {11};
\node[circle,minimum size=0.8cm,draw=black,fill=white] at (8.,2.0) {11};
\node[circle,minimum size=0.8cm,draw=black,fill=white] at (8.,1.0) {11};
\node[circle,minimum size=0.8cm,draw=red,fill=lightgray] at (8.,0.0) {6};
\node[circle,minimum size=0.8cm,draw=black,fill=white] at (8.,-1.0) {11};
\node[circle,minimum size=0.8cm,draw=black,fill=white] at (8.,-2.0) {11};
\node[circle,minimum size=0.8cm,draw=black,fill=white] at (9.,1.0) {11};
\node[circle,minimum size=0.8cm,draw=black,fill=lightgray] at (9.,0.0) {3};
\node[circle,minimum size=0.8cm,draw=black,fill=white] at (9.,-1.0) {11};
\node[circle,minimum size=0.8cm,draw=black,fill=lightgray] at (10.,0.0) {6};
\end{tikzpicture}
\\
\vspace{-0.2cm}
\(-N_{11},+N_{3a}\)
\vspace{0.2cm}
\end{center}
The same move but with as a consequence a formation of a structure \(\textbf C\)
has \textit{mutatis mutandis} probability \(pN_6/2\), and we lose 2 states (11)
and gain 1 (3b) and 1 (10c):
\begin{center}
\vspace{0.2cm}
\(P=pN_{6}/2\)
\\
\begin{tikzpicture}
\draw[thick][red]  (3.,0) -- (3,1);
\draw[thick][red]  (8.,0) -- (9,0);
\draw[thick][red]  (9.,0) -- (9,1);
\node[circle,minimum size=0.8cm,draw=black,fill=white] at (0.,0.0) {11};
\node[circle,minimum size=0.8cm,draw=black,fill=white] at (1.,1.0) {11};
\node[circle,minimum size=0.8cm,draw=black,fill=white] at (1.,0.0) {11};
\node[circle,minimum size=0.8cm,draw=black,fill=white] at (1.,-1.0) {11};
\node[circle,minimum size=0.8cm,draw=black,fill=white] at (2.,2.0) {11};
\node[circle,minimum size=0.8cm,draw=black,fill=white] at (2.,1.0) {11};
\node[circle,minimum size=0.8cm,draw=red,fill=white] at (2.,0.0) {11};
\node[circle,minimum size=0.8cm,draw=black,fill=white] at (2.,-1.0) {11};
\node[circle,minimum size=0.8cm,draw=black,fill=white] at (2.,-2.0) {11};
\node[circle,minimum size=0.8cm,draw=black,fill=lightgray] at (3.,1.0) {6};
\node[circle,minimum size=0.8cm,draw=black,fill=lightgray] at (3.,0.0) {6};
\node[circle,minimum size=0.8cm,draw=black,fill=white] at (3.,-1.0) {11};
\node[circle,minimum size=0.8cm,draw=black,fill=white] at (4.,0.0) {11};
\draw [very thick] [ ->] (4.6,0.0) -- (5.4,0.0);
\node[circle,minimum size=0.8cm,draw=black,fill=white] at (6.,0.0) {11};
\node[circle,minimum size=0.8cm,draw=black,fill=white] at (7.,1.0) {11};
\node[circle,minimum size=0.8cm,draw=black,fill=white] at (7.,0.0) {11};
\node[circle,minimum size=0.8cm,draw=black,fill=white] at (7.,-1.0) {11};
\node[circle,minimum size=0.8cm,draw=black,fill=white] at (8.,2.0) {11};
\node[circle,minimum size=0.8cm,draw=black,fill=white] at (8.,1.0) {10};
\node[circle,minimum size=0.8cm,draw=red,fill=lightgray] at (8.,0.0) {6};
\node[circle,minimum size=0.8cm,draw=black,fill=white] at (8.,-1.0) {11};
\node[circle,minimum size=0.8cm,draw=black,fill=white] at (8.,-2.0) {11};
\node[circle,minimum size=0.8cm,draw=black,fill=lightgray] at (9.,1.0) {6};
\node[circle,minimum size=0.8cm,draw=black,fill=lightgray] at (9.,0.0) {3};
\node[circle,minimum size=0.8cm,draw=black,fill=white] at (9.,-1.0) {11};
\node[circle,minimum size=0.8cm,draw=black,fill=white] at (10.,0.0) {11};
\end{tikzpicture}
\\
\vspace{-0.2cm}
\(-2N_{11},+N_{3b},+N_{10c}\)
\end{center}


A site in a state (11) that is far from any structures and
flips to another \(q\) value but remains in the state (11)
can, with probability \(N_{11}4/(4e^{\beta}+q-4)\)
assume the same colour of one of its next to nearest neighbours thus forming an 
\(\textbf E\) or, again with probability, \(N_{11}4/(4e^{\beta}+q-4)\) form an \(\textbf F\) structure.
We have, respectively, 
\begin{center}
\vspace{0.2cm}
\(P=N_{11}(4/(4e^{\beta}+q-4))\)
\nopagebreak
\\
\begin{tikzpicture}
\node[circle,minimum size=0.8cm,draw=black,fill=white] at (0.,0.0) {11};
\node[circle,minimum size=0.8cm,draw=black,fill=white] at (1.,1.0) {11};
\node[circle,minimum size=0.8cm,draw=black,fill=white] at (1.,0.0) {11};
\node[circle,minimum size=0.8cm,draw=black,fill=white] at (1.,-1.0) {11};
\node[circle,minimum size=0.8cm,draw=black,fill=white] at (2.,2.0) {11};
\node[circle,minimum size=0.8cm,draw=black,fill=white] at (2.,1.0) {11};
\node[circle,minimum size=0.8cm,draw=red,fill=white] at (2.,0.0) {11};
\node[circle,minimum size=0.8cm,draw=black,fill=white] at (2.,-1.0) {11};
\node[circle,minimum size=0.8cm,draw=black,fill=white] at (2.,-2.0) {11};
\node[circle,minimum size=0.8cm,draw=black,fill=white] at (3.,1.0) {11};
\node[circle,minimum size=0.8cm,draw=black,fill=white] at (3.,0.0) {11};
\node[circle,minimum size=0.8cm,draw=black,fill=white] at (3.,-1.0) {11};
\node[circle,minimum size=0.8cm,draw=black,fill=lightgray] at (4.,0.0) {11};
\draw [very thick] [ ->] (4.6,0.0) -- (5.4,0.0);
\node[circle,minimum size=0.8cm,draw=black,fill=white] at (6.,0.0) {11};
\node[circle,minimum size=0.8cm,draw=black,fill=white] at (7.,1.0) {11};
\node[circle,minimum size=0.8cm,draw=black,fill=white] at (7.,0.0) {11};
\node[circle,minimum size=0.8cm,draw=black,fill=white] at (7.,-1.0) {11};
\node[circle,minimum size=0.8cm,draw=black,fill=white] at (8.,2.0) {11};
\node[circle,minimum size=0.8cm,draw=black,fill=white] at (8.,1.0) {11};
\node[circle,minimum size=0.8cm,draw=red,fill=lightgray] at (8.,0.0) {11};
\node[circle,minimum size=0.8cm,draw=black,fill=white] at (8.,-1.0) {11};
\node[circle,minimum size=0.8cm,draw=black,fill=white] at (8.,-2.0) {11};
\node[circle,minimum size=0.8cm,draw=black,fill=white] at (9.,1.0) {11};
\node[circle,minimum size=0.8cm,draw=black,fill=white] at (9.,0.0) {10};
\node[circle,minimum size=0.8cm,draw=black,fill=white] at (9.,-1.0) {11};
\node[circle,minimum size=0.8cm,draw=black,fill=lightgray] at (10.,0.0) {11};
\end{tikzpicture}
\\
\vspace{-0.2cm}
\(-N_{11},+N_{10a}\)
\end{center}
and 
\begin{center}
\(P=N_{11}(4/(4e^{\beta}+q-4))\)
\\
\begin{tikzpicture}
\node[circle,minimum size=0.8cm,draw=black,fill=white] at (0.,0.0) {11};
\node[circle,minimum size=0.8cm,draw=black,fill=white] at (1.,1.0) {11};
\node[circle,minimum size=0.8cm,draw=black,fill=white] at (1.,0.0) {11};
\node[circle,minimum size=0.8cm,draw=black,fill=white] at (1.,-1.0) {11};
\node[circle,minimum size=0.8cm,draw=black,fill=white] at (2.,2.0) {11};
\node[circle,minimum size=0.8cm,draw=black,fill=white] at (2.,1.0) {11};
\node[circle,minimum size=0.8cm,draw=red,fill=white] at (2.,0.0) {11};
\node[circle,minimum size=0.8cm,draw=black,fill=white] at (2.,-1.0) {11};
\node[circle,minimum size=0.8cm,draw=black,fill=white] at (2.,-2.0) {11};
\node[circle,minimum size=0.8cm,draw=black,fill=lightgray] at (3.,1.0) {11};
\node[circle,minimum size=0.8cm,draw=black,fill=white] at (3.,0.0) {11};
\node[circle,minimum size=0.8cm,draw=black,fill=white] at (3.,-1.0) {11};
\node[circle,minimum size=0.8cm,draw=black,fill=white] at (4.,0.0) {11};
\draw [very thick] [ ->] (4.6,0.0) -- (5.4,0.0);
\node[circle,minimum size=0.8cm,draw=black,fill=white] at (6.,0.0) {11};
\node[circle,minimum size=0.8cm,draw=black,fill=white] at (7.,1.0) {11};
\node[circle,minimum size=0.8cm,draw=black,fill=white] at (7.,0.0) {11};
\node[circle,minimum size=0.8cm,draw=black,fill=white] at (7.,-1.0) {11};
\node[circle,minimum size=0.8cm,draw=black,fill=white] at (8.,2.0) {11};
\node[circle,minimum size=0.8cm,draw=black,fill=white] at (8.,1.0) {10};
\node[circle,minimum size=0.8cm,draw=red,fill=lightgray] at (8.,0.0) {11};
\node[circle,minimum size=0.8cm,draw=black,fill=white] at (8.,-1.0) {11};
\node[circle,minimum size=0.8cm,draw=black,fill=white] at (8.,-2.0) {11};
\node[circle,minimum size=0.8cm,draw=black,fill=lightgray] at (9.,1.0) {11};
\node[circle,minimum size=0.8cm,draw=black,fill=white] at (9.,0.0) {10};
\node[circle,minimum size=0.8cm,draw=black,fill=white] at (9.,-1.0) {11};
\node[circle,minimum size=0.8cm,draw=black,fill=white] at (10.,0.0) {11};
\end{tikzpicture}
\\
\vspace{-0.25cm}
\(-2N_{11},+2N_{10b}\)
\vspace{0.25cm}
\end{center}

Picking one of the two gray sites which are part of an \(\textbf E\) structure 
has probability \(2P(\textbf E)=2N_{10a}\). The probability for it to change colour 
but stay in a state (11) is \(P_{11\rightarrow11}=1-p\).
Thus, the following move
\begin{center}
\(P=2N_{10a}\)
\nopagebreak
\\
\vspace{-0.2cm}
\begin{tikzpicture}
\node[circle,minimum size=0.8cm,draw=black,fill=white] at (0.,0.0) {11};
\node[circle,minimum size=0.8cm,draw=black,fill=white] at (1.,1.0) {11};
\node[circle,minimum size=0.8cm,draw=black,fill=white] at (1.,0.0) {11};
\node[circle,minimum size=0.8cm,draw=black,fill=white] at (1.,-1.0) {11};
\node[circle,minimum size=0.8cm,draw=black,fill=white] at (2.,2.0) {11};
\node[circle,minimum size=0.8cm,draw=black,fill=white] at (2.,1.0) {11};
\node[circle,minimum size=0.8cm,draw=red,fill=lightgray] at (2.,0.0) {11};
\node[circle,minimum size=0.8cm,draw=black,fill=white] at (2.,-1.0) {11};
\node[circle,minimum size=0.8cm,draw=black,fill=white] at (2.,-2.0) {11};
\node[circle,minimum size=0.8cm,draw=black,fill=white] at (3.,1.0) {11};
\node[circle,minimum size=0.8cm,draw=black,fill=white] at (3.,0.0) {10};
\node[circle,minimum size=0.8cm,draw=black,fill=white] at (3.,-1.0) {11};
\node[circle,minimum size=0.8cm,draw=black,fill=lightgray] at (4.,0.0) {11};
\draw [very thick] [ ->] (4.6,0.0) -- (5.4,0.0);
\node[circle,minimum size=0.8cm,draw=black,fill=white] at (6.,0.0) {11};
\node[circle,minimum size=0.8cm,draw=black,fill=white] at (7.,1.0) {11};
\node[circle,minimum size=0.8cm,draw=black,fill=white] at (7.,0.0) {11};
\node[circle,minimum size=0.8cm,draw=black,fill=white] at (7.,-1.0) {11};
\node[circle,minimum size=0.8cm,draw=black,fill=white] at (8.,2.0) {11};
\node[circle,minimum size=0.8cm,draw=black,fill=white] at (8.,1.0) {11};
\node[circle,minimum size=0.8cm,draw=red,fill=white] at (8.,0.0) {11};
\node[circle,minimum size=0.8cm,draw=black,fill=white] at (8.,-1.0) {11};
\node[circle,minimum size=0.8cm,draw=black,fill=white] at (8.,-2.0) {11};
\node[circle,minimum size=0.8cm,draw=black,fill=white] at (9.,1.0) {11};
\node[circle,minimum size=0.8cm,draw=black,fill=white] at (9.,0.0) {11};
\node[circle,minimum size=0.8cm,draw=black,fill=white] at (9.,-1.0) {11};
\node[circle,minimum size=0.8cm,draw=black,fill=lightgray] at (10.,0.0) {11};
\end{tikzpicture}
\\
\(-N_{10a},+N_{11}\)
\vspace{0.5cm}
\nopagebreak
\end{center}
occurs with probability
\(2N_{10a}\) and causes a loss of a (10a) and a gain of an (11)

Similarly, there are two gray (11) which are part of a structure \(\textbf F\).
Thus with probability \(N_{10b}\) the following move cause a loss of 2 (10b) and the gain of 
2 (11)
\begin{center}
\vspace{0.5cm}
\(P=N_{10b}\)
\\
\nopagebreak
\vspace{0.25cm}
\begin{tikzpicture}
\node[circle,minimum size=0.8cm,draw=black,fill=white] at (0.,0.0) {11};
\node[circle,minimum size=0.8cm,draw=black,fill=white] at (1.,1.0) {11};
\node[circle,minimum size=0.8cm,draw=black,fill=white] at (1.,0.0) {11};
\node[circle,minimum size=0.8cm,draw=black,fill=white] at (1.,-1.0) {11};
\node[circle,minimum size=0.8cm,draw=black,fill=white] at (2.,2.0) {11};
\node[circle,minimum size=0.8cm,draw=black,fill=white] at (2.,1.0) {10};
\node[circle,minimum size=0.8cm,draw=red,fill=lightgray] at (2.,0.0) {11};
\node[circle,minimum size=0.8cm,draw=black,fill=white] at (2.,-1.0) {11};
\node[circle,minimum size=0.8cm,draw=black,fill=white] at (2.,-2.0) {11};
\node[circle,minimum size=0.8cm,draw=black,fill=lightgray] at (3.,1.0) {11};
\node[circle,minimum size=0.8cm,draw=black,fill=white] at (3.,0.0) {10};
\node[circle,minimum size=0.8cm,draw=black,fill=white] at (3.,-1.0) {11};
\node[circle,minimum size=0.8cm,draw=black,fill=white] at (4.,0.0) {11};
\draw [very thick] [ ->] (4.6,0.0) -- (5.4,0.0);
\node[circle,minimum size=0.8cm,draw=black,fill=white] at (6.,0.0) {11};
\node[circle,minimum size=0.8cm,draw=black,fill=white] at (7.,1.0) {11};
\node[circle,minimum size=0.8cm,draw=black,fill=white] at (7.,0.0) {11};
\node[circle,minimum size=0.8cm,draw=black,fill=white] at (7.,-1.0) {11};
\node[circle,minimum size=0.8cm,draw=black,fill=white] at (8.,2.0) {11};
\node[circle,minimum size=0.8cm,draw=black,fill=white] at (8.,1.0) {11};
\node[circle,minimum size=0.8cm,draw=red,fill=white] at (8.,0.0) {11};
\node[circle,minimum size=0.8cm,draw=black,fill=white] at (8.,-1.0) {11};
\node[circle,minimum size=0.8cm,draw=black,fill=white] at (8.,-2.0) {11};
\node[circle,minimum size=0.8cm,draw=black,fill=lightgray] at (9.,1.0) {11};
\node[circle,minimum size=0.8cm,draw=black,fill=white] at (9.,0.0) {11};
\node[circle,minimum size=0.8cm,draw=black,fill=white] at (9.,-1.0) {11};
\node[circle,minimum size=0.8cm,draw=black,fill=white] at (10.,0.0) {11};
\end{tikzpicture}
\\
\(-2N_{10b},+2N_{11}\)
\vspace{0.5cm}
\end{center}

Now we consider the cases when the starting state is a (6).
The probability of picking a (6) which is part of a structure \(\textbf A\)
is \(2P(\textbf A)=N_{6}-2N_{3a}-2N_{3b}\) and it turns to a (11)
with probability \(1-p\). So, to the second order in \(p^2\),
with probability \(N_{6}(1-p)-2(N_{3a}+N_{3b})\), 2 (6) disappears
and 2 (11) appears
\begin{center}
\vspace{0.5cm}
\(P=N_{6}(1-p)-2(N_{3a}+N_{3b})\)
\nopagebreak
\\
\vspace{0.25cm}
\begin{tikzpicture}
\draw[thick][red]  (2.,0) -- (3,0);
\node[circle,minimum size=0.8cm,draw=black,fill=white] at (0.,0.0) {11};
\node[circle,minimum size=0.8cm,draw=black,fill=white] at (1.,1.0) {11};
\node[circle,minimum size=0.8cm,draw=black,fill=white] at (1.,0.0) {11};
\node[circle,minimum size=0.8cm,draw=black,fill=white] at (1.,-1.0) {11};
\node[circle,minimum size=0.8cm,draw=black,fill=white] at (2.,2.0) {11};
\node[circle,minimum size=0.8cm,draw=black,fill=white] at (2.,1.0) {11};
\node[circle,minimum size=0.8cm,draw=red,fill=lightgray] at (2.,0.0) {6};
\node[circle,minimum size=0.8cm,draw=black,fill=white] at (2.,-1.0) {11};
\node[circle,minimum size=0.8cm,draw=black,fill=white] at (2.,-2.0) {11};
\node[circle,minimum size=0.8cm,draw=black,fill=white] at (3.,1.0) {11};
\node[circle,minimum size=0.8cm,draw=black,fill=lightgray] at (3.,0.0) {6};
\node[circle,minimum size=0.8cm,draw=black,fill=white] at (3.,-1.0) {11};
\node[circle,minimum size=0.8cm,draw=black,fill=white] at (4.,0.0) {11};
\draw [very thick] [ ->] (4.6,0.0) -- (5.4,0.0);
\node[circle,minimum size=0.8cm,draw=black,fill=white] at (6.,0.0) {11};
\node[circle,minimum size=0.8cm,draw=black,fill=white] at (7.,1.0) {11};
\node[circle,minimum size=0.8cm,draw=black,fill=white] at (7.,0.0) {11};
\node[circle,minimum size=0.8cm,draw=black,fill=white] at (7.,-1.0) {11};
\node[circle,minimum size=0.8cm,draw=black,fill=white] at (8.,2.0) {11};
\node[circle,minimum size=0.8cm,draw=black,fill=white] at (8.,1.0) {11};
\node[circle,minimum size=0.8cm,draw=red,fill=white] at (8.,0.0) {11};
\node[circle,minimum size=0.8cm,draw=black,fill=white] at (8.,-1.0) {11};
\node[circle,minimum size=0.8cm,draw=black,fill=white] at (8.,-2.0) {11};
\node[circle,minimum size=0.8cm,draw=black,fill=white] at (9.,1.0) {11};
\node[circle,minimum size=0.8cm,draw=black,fill=lightgray] at (9.,0.0) {11};
\node[circle,minimum size=0.8cm,draw=black,fill=white] at (9.,-1.0) {11};
\node[circle,minimum size=0.8cm,draw=black,fill=white] at (10.,0.0) {11};
\end{tikzpicture}
\\
\(-2N_{6},+2N_{11}\)
\vspace{0.5cm}
\end{center}

Considering instead picking one of the two (6) which are part of a structure 
\(\textbf B\) or \(\textbf C\), the transition to a (11)
leads respectively with probability \(2N_{3a}\) to a loss of 1 (3a) and 
a gain of a (11)
\begin{center}
\vspace{0.5cm}
\(P=2N_{3a}\)
\nopagebreak
\\
\vspace{0.25cm}
\begin{tikzpicture}
\draw[thick][red]  (2.,0) -- (3,0);
\draw[thick][red]  (3.,0) -- (4,0);
\draw[thick][red]  (9.,0) -- (10,0);
\node[circle,minimum size=0.8cm,draw=black,fill=white] at (0.,0.0) {11};
\node[circle,minimum size=0.8cm,draw=black,fill=white] at (1.,1.0) {11};
\node[circle,minimum size=0.8cm,draw=black,fill=white] at (1.,0.0) {11};
\node[circle,minimum size=0.8cm,draw=black,fill=white] at (1.,-1.0) {11};
\node[circle,minimum size=0.8cm,draw=black,fill=white] at (2.,2.0) {11};
\node[circle,minimum size=0.8cm,draw=black,fill=white] at (2.,1.0) {11};
\node[circle,minimum size=0.8cm,draw=red,fill=lightgray] at (2.,0.0) {6};
\node[circle,minimum size=0.8cm,draw=black,fill=white] at (2.,-1.0) {11};
\node[circle,minimum size=0.8cm,draw=black,fill=white] at (2.,-2.0) {11};
\node[circle,minimum size=0.8cm,draw=black,fill=white] at (3.,1.0) {11};
\node[circle,minimum size=0.8cm,draw=black,fill=lightgray] at (3.,0.0) {3};
\node[circle,minimum size=0.8cm,draw=black,fill=white] at (3.,-1.0) {11};
\node[circle,minimum size=0.8cm,draw=black,fill=lightgray] at (4.,0.0) {6};
\draw [very thick] [ ->] (4.6,0.0) -- (5.4,0.0);
\node[circle,minimum size=0.8cm,draw=black,fill=white] at (6.,0.0) {11};
\node[circle,minimum size=0.8cm,draw=black,fill=white] at (7.,1.0) {11};
\node[circle,minimum size=0.8cm,draw=black,fill=white] at (7.,0.0) {11};
\node[circle,minimum size=0.8cm,draw=black,fill=white] at (7.,-1.0) {11};
\node[circle,minimum size=0.8cm,draw=black,fill=white] at (8.,2.0) {11};
\node[circle,minimum size=0.8cm,draw=black,fill=white] at (8.,1.0) {11};
\node[circle,minimum size=0.8cm,draw=red,fill=white] at (8.,0.0) {11};
\node[circle,minimum size=0.8cm,draw=black,fill=white] at (8.,-1.0) {11};
\node[circle,minimum size=0.8cm,draw=black,fill=white] at (8.,-2.0) {11};
\node[circle,minimum size=0.8cm,draw=black,fill=white] at (9.,1.0) {11};
\node[circle,minimum size=0.8cm,draw=black,fill=lightgray] at (9.,0.0) {6};
\node[circle,minimum size=0.8cm,draw=black,fill=white] at (9.,-1.0) {11};
\node[circle,minimum size=0.8cm,draw=black,fill=lightgray] at (10.,0.0) {6};
\end{tikzpicture}
\\
\(-N_{3a},+N_{11}\)
\vspace{0.5cm}
\end{center}
and with probability \(2N_{3b}\) to the distruction of 1 (3b) and 1 (10c)
and the creation of 2 (11)
\begin{center}
\vspace{0.5cm}
\(P=2N_{3b}\)
\nopagebreak
\\
\vspace{0.25cm}
\begin{tikzpicture}
\draw[thick][red]  (2.,0) -- (3,0);
\draw[thick][red]  (3.,0) -- (3,1);
\draw[thick][red]  (9.,0) -- (9,1);
\node[circle,minimum size=0.8cm,draw=black,fill=white] at (0.,0.0) {11};
\node[circle,minimum size=0.8cm,draw=black,fill=white] at (1.,1.0) {11};
\node[circle,minimum size=0.8cm,draw=black,fill=white] at (1.,0.0) {11};
\node[circle,minimum size=0.8cm,draw=black,fill=white] at (1.,-1.0) {11};
\node[circle,minimum size=0.8cm,draw=black,fill=white] at (2.,2.0) {11};
\node[circle,minimum size=0.8cm,draw=black,fill=white] at (2.,1.0) {10};
\node[circle,minimum size=0.8cm,draw=red,fill=lightgray] at (2.,0.0) {6};
\node[circle,minimum size=0.8cm,draw=black,fill=white] at (2.,-1.0) {11};
\node[circle,minimum size=0.8cm,draw=black,fill=white] at (2.,-2.0) {11};
\node[circle,minimum size=0.8cm,draw=black,fill=lightgray] at (3.,1.0) {6};
\node[circle,minimum size=0.8cm,draw=black,fill=lightgray] at (3.,0.0) {3};
\node[circle,minimum size=0.8cm,draw=black,fill=white] at (3.,-1.0) {11};
\node[circle,minimum size=0.8cm,draw=black,fill=white] at (4.,0.0) {11};
\draw [very thick] [ ->] (4.6,0.0) -- (5.4,0.0);
\node[circle,minimum size=0.8cm,draw=black,fill=white] at (6.,0.0) {11};
\node[circle,minimum size=0.8cm,draw=black,fill=white] at (7.,1.0) {11};
\node[circle,minimum size=0.8cm,draw=black,fill=white] at (7.,0.0) {11};
\node[circle,minimum size=0.8cm,draw=black,fill=white] at (7.,-1.0) {11};
\node[circle,minimum size=0.8cm,draw=black,fill=white] at (8.,2.0) {11};
\node[circle,minimum size=0.8cm,draw=black,fill=white] at (8.,1.0) {11};
\node[circle,minimum size=0.8cm,draw=red,fill=white] at (8.,0.0) {11};
\node[circle,minimum size=0.8cm,draw=black,fill=white] at (8.,-1.0) {11};
\node[circle,minimum size=0.8cm,draw=black,fill=white] at (8.,-2.0) {11};
\node[circle,minimum size=0.8cm,draw=black,fill=lightgray] at (9.,1.0) {6};
\node[circle,minimum size=0.8cm,draw=black,fill=lightgray] at (9.,0.0) {6};
\node[circle,minimum size=0.8cm,draw=black,fill=white] at (9.,-1.0) {11};
\node[circle,minimum size=0.8cm,draw=black,fill=white] at (10.,0.0) {11};
\end{tikzpicture}
\\
\(-N_{3b},-N_{10c},+2N_{11}\)
\vspace{0.5cm}
\end{center}

Now we consider all the moves involving as starting sites a (3) or a (10)
of all the possible kinds. This states, which can be picked 
with probability proportional to \(p^2\) can turn one into the other
with probabilities \(P_{3\rightarrow10}\sim1/2\) and \(P_{10\rightarrow3}\sim1/2\)
for \(T\simeq T_c\). 
Consider picking a (3a), this happens with probability \(N_{3a}\),
if it turns into a (10) (it happens with probability \(P_{3\rightarrow10}\)) it cause the loss of 1 (3a) and 2 (6) and the gain of 1 (10a) and 2 (11)
\begin{center}
\vspace{0.5cm}
\(P=N_{3a}P_{3\rightarrow10}\)
\nopagebreak
\\
\vspace{0.25cm}
\begin{tikzpicture}
\draw[thick][red]  (1.,0) -- (2,0);
\draw[thick][red]  (2.,0) -- (3,0);
\node[circle,minimum size=0.8cm,draw=black,fill=white] at (0.,0.0) {11};
\node[circle,minimum size=0.8cm,draw=black,fill=white] at (1.,1.0) {11};
\node[circle,minimum size=0.8cm,draw=black,fill=lightgray] at (1.,0.0) {6};
\node[circle,minimum size=0.8cm,draw=black,fill=white] at (1.,-1.0) {11};
\node[circle,minimum size=0.8cm,draw=black,fill=white] at (2.,2.0) {11};
\node[circle,minimum size=0.8cm,draw=black,fill=white] at (2.,1.0) {11};
\node[circle,minimum size=0.8cm,draw=red,fill=lightgray] at (2.,0.0) {3};
\node[circle,minimum size=0.8cm,draw=black,fill=white] at (2.,-1.0) {11};
\node[circle,minimum size=0.8cm,draw=black,fill=white] at (2.,-2.0) {11};
\node[circle,minimum size=0.8cm,draw=black,fill=white] at (3.,1.0) {11};
\node[circle,minimum size=0.8cm,draw=black,fill=lightgray] at (3.,0.0) {6};
\node[circle,minimum size=0.8cm,draw=black,fill=white] at (3.,-1.0) {11};
\node[circle,minimum size=0.8cm,draw=black,fill=white] at (4.,0.0) {11};
\draw [very thick] [ ->] (4.6,0.0) -- (5.4,0.0);
\node[circle,minimum size=0.8cm,draw=black,fill=white] at (6.,0.0) {11};
\node[circle,minimum size=0.8cm,draw=black,fill=white] at (7.,1.0) {11};
\node[circle,minimum size=0.8cm,draw=black,fill=lightgray] at (7.,0.0) {11};
\node[circle,minimum size=0.8cm,draw=black,fill=white] at (7.,-1.0) {11};
\node[circle,minimum size=0.8cm,draw=black,fill=white] at (8.,2.0) {11};
\node[circle,minimum size=0.8cm,draw=black,fill=white] at (8.,1.0) {11};
\node[circle,minimum size=0.8cm,draw=red,fill=white] at (8.,0.0) {10};
\node[circle,minimum size=0.8cm,draw=black,fill=white] at (8.,-1.0) {11};
\node[circle,minimum size=0.8cm,draw=black,fill=white] at (8.,-2.0) {11};
\node[circle,minimum size=0.8cm,draw=black,fill=white] at (9.,1.0) {11};
\node[circle,minimum size=0.8cm,draw=black,fill=lightgray] at (9.,0.0) {11};
\node[circle,minimum size=0.8cm,draw=black,fill=white] at (9.,-1.0) {11};
\node[circle,minimum size=0.8cm,draw=black,fill=white] at (10.,0.0) {11};
\end{tikzpicture}
\\
\(-N_{3a},-2N_{6},+N_{10a},+2N_{11}\)
\vspace{0.5cm}
\end{center}

The inverse is
\begin{center}
\(P=N_{10a}P_{10\rightarrow3}\)
\nopagebreak
\\
\vspace{0.25cm}
\begin{tikzpicture}
\draw[thick][red]  (7.,0) -- (8,0);
\draw[thick][red]  (8.,0) -- (9,0);
\node[circle,minimum size=0.8cm,draw=black,fill=white] at (0.,0.0) {11};
\node[circle,minimum size=0.8cm,draw=black,fill=white] at (1.,1.0) {11};
\node[circle,minimum size=0.8cm,draw=black,fill=lightgray] at (1.,0.0) {11};
\node[circle,minimum size=0.8cm,draw=black,fill=white] at (1.,-1.0) {11};
\node[circle,minimum size=0.8cm,draw=black,fill=white] at (2.,2.0) {11};
\node[circle,minimum size=0.8cm,draw=black,fill=white] at (2.,1.0) {11};
\node[circle,minimum size=0.8cm,draw=red,fill=white] at (2.,0.0) {10};
\node[circle,minimum size=0.8cm,draw=black,fill=white] at (2.,-1.0) {11};
\node[circle,minimum size=0.8cm,draw=black,fill=white] at (2.,-2.0) {11};
\node[circle,minimum size=0.8cm,draw=black,fill=white] at (3.,1.0) {11};
\node[circle,minimum size=0.8cm,draw=black,fill=lightgray] at (3.,0.0) {11};
\node[circle,minimum size=0.8cm,draw=black,fill=white] at (3.,-1.0) {11};
\node[circle,minimum size=0.8cm,draw=black,fill=white] at (4.,0.0) {11};
\draw [very thick] [ ->] (4.6,0.0) -- (5.4,0.0);
\node[circle,minimum size=0.8cm,draw=black,fill=white] at (6.,0.0) {11};
\node[circle,minimum size=0.8cm,draw=black,fill=white] at (7.,1.0) {11};
\node[circle,minimum size=0.8cm,draw=black,fill=lightgray] at (7.,0.0) {6};
\node[circle,minimum size=0.8cm,draw=black,fill=white] at (7.,-1.0) {11};
\node[circle,minimum size=0.8cm,draw=black,fill=white] at (8.,2.0) {11};
\node[circle,minimum size=0.8cm,draw=black,fill=white] at (8.,1.0) {11};
\node[circle,minimum size=0.8cm,draw=red,fill=lightgray] at (8.,0.0) {3};
\node[circle,minimum size=0.8cm,draw=black,fill=white] at (8.,-1.0) {11};
\node[circle,minimum size=0.8cm,draw=black,fill=white] at (8.,-2.0) {11};
\node[circle,minimum size=0.8cm,draw=black,fill=white] at (9.,1.0) {11};
\node[circle,minimum size=0.8cm,draw=black,fill=lightgray] at (9.,0.0) {6};
\node[circle,minimum size=0.8cm,draw=black,fill=white] at (9.,-1.0) {11};
\node[circle,minimum size=0.8cm,draw=black,fill=white] at (10.,0.0) {11};
\end{tikzpicture}
\\
\(-N_{10a},-2N_{11},+N_{3a},+2N_{6}\)
\vspace{0.5cm}
\end{center}

If we pick a (3b), the probability of the move is 
\(N_{3b}P_{3\rightarrow10}\) and cause 
the destruction of 1 (3b), 1 (10c) and 2 (6) while creates
2 (10b) and 2 (11). We have
\begin{center}
\vspace{0.5cm}
\(P=N_{3b}P_{3\rightarrow10}\)
\nopagebreak
\\
\vspace{0.25cm}
\begin{tikzpicture}
\draw[thick][red]  (2.,0) -- (2,1);
\draw[thick][red]  (2.,0) -- (3,0);
\node[circle,minimum size=0.8cm,draw=black,fill=white] at (0.,0.0) {11};
\node[circle,minimum size=0.8cm,draw=black,fill=white] at (1.,1.0) {11};
\node[circle,minimum size=0.8cm,draw=black,fill=white] at (1.,0.0) {11};
\node[circle,minimum size=0.8cm,draw=black,fill=white] at (1.,-1.0) {11};
\node[circle,minimum size=0.8cm,draw=black,fill=white] at (2.,2.0) {11};
\node[circle,minimum size=0.8cm,draw=black,fill=lightgray] at (2.,1.0) {6};
\node[circle,minimum size=0.8cm,draw=red,fill=lightgray] at (2.,0.0) {3};
\node[circle,minimum size=0.8cm,draw=black,fill=white] at (2.,-1.0) {11};
\node[circle,minimum size=0.8cm,draw=black,fill=white] at (2.,-2.0) {11};
\node[circle,minimum size=0.8cm,draw=black,fill=white] at (3.,1.0) {10};
\node[circle,minimum size=0.8cm,draw=black,fill=lightgray] at (3.,0.0) {6};
\node[circle,minimum size=0.8cm,draw=black,fill=white] at (3.,-1.0) {11};
\node[circle,minimum size=0.8cm,draw=black,fill=white] at (4.,0.0) {11};
\draw [very thick] [ ->] (4.6,0.0) -- (5.4,0.0);
\node[circle,minimum size=0.8cm,draw=black,fill=white] at (6.,0.0) {11};
\node[circle,minimum size=0.8cm,draw=black,fill=white] at (7.,1.0) {11};
\node[circle,minimum size=0.8cm,draw=black,fill=white] at (7.,0.0) {11};
\node[circle,minimum size=0.8cm,draw=black,fill=white] at (7.,-1.0) {11};
\node[circle,minimum size=0.8cm,draw=black,fill=white] at (8.,2.0) {11};
\node[circle,minimum size=0.8cm,draw=black,fill=lightgray] at (8.,1.0) {11};
\node[circle,minimum size=0.8cm,draw=red,fill=white] at (8.,0.0) {10};
\node[circle,minimum size=0.8cm,draw=black,fill=white] at (8.,-1.0) {11};
\node[circle,minimum size=0.8cm,draw=black,fill=white] at (8.,-2.0) {11};
\node[circle,minimum size=0.8cm,draw=black,fill=white] at (9.,1.0) {10};
\node[circle,minimum size=0.8cm,draw=black,fill=lightgray] at (9.,0.0) {11};
\node[circle,minimum size=0.8cm,draw=black,fill=white] at (9.,-1.0) {11};
\node[circle,minimum size=0.8cm,draw=black,fill=white] at (10.,0.0) {11};
\end{tikzpicture}
\nopagebreak
\\
\vspace{0.125cm}
\(-N_{3b},-N_{10c},-2N_{6},+2N_{10b},+2N_{11}\)
\end{center}

\noindent
and the opposite move with 
\begin{center}
\vspace{-0.25cm}
\(P=N_{10b}P_{10\rightarrow3}\)
\nopagebreak
\\
\vspace{0.25cm}
\begin{tikzpicture}
\draw[thick][red]  (8.,0) -- (8,1);
\draw[thick][red]  (8.,0) -- (9,0);
\node[circle,minimum size=0.8cm,draw=black,fill=white] at (0.,0.0) {11};
\node[circle,minimum size=0.8cm,draw=black,fill=white] at (1.,1.0) {11};
\node[circle,minimum size=0.8cm,draw=black,fill=white] at (1.,0.0) {11};
\node[circle,minimum size=0.8cm,draw=black,fill=white] at (1.,-1.0) {11};
\node[circle,minimum size=0.8cm,draw=black,fill=white] at (2.,2.0) {11};
\node[circle,minimum size=0.8cm,draw=black,fill=lightgray] at (2.,1.0) {11};
\node[circle,minimum size=0.8cm,draw=red,fill=white] at (2.,0.0) {10};
\node[circle,minimum size=0.8cm,draw=black,fill=white] at (2.,-1.0) {11};
\node[circle,minimum size=0.8cm,draw=black,fill=white] at (2.,-2.0) {11};
\node[circle,minimum size=0.8cm,draw=black,fill=white] at (3.,1.0) {10};
\node[circle,minimum size=0.8cm,draw=black,fill=lightgray] at (3.,0.0) {11};
\node[circle,minimum size=0.8cm,draw=black,fill=white] at (3.,-1.0) {11};
\node[circle,minimum size=0.8cm,draw=black,fill=white] at (4.,0.0) {11};
\draw [very thick] [ ->] (4.6,0.0) -- (5.4,0.0);
\node[circle,minimum size=0.8cm,draw=black,fill=white] at (6.,0.0) {11};
\node[circle,minimum size=0.8cm,draw=black,fill=white] at (7.,1.0) {11};
\node[circle,minimum size=0.8cm,draw=black,fill=white] at (7.,0.0) {11};
\node[circle,minimum size=0.8cm,draw=black,fill=white] at (7.,-1.0) {11};
\node[circle,minimum size=0.8cm,draw=black,fill=white] at (8.,2.0) {11};
\node[circle,minimum size=0.8cm,draw=black,fill=lightgray] at (8.,1.0) {6};
\node[circle,minimum size=0.8cm,draw=red,fill=lightgray] at (8.,0.0) {3};
\node[circle,minimum size=0.8cm,draw=black,fill=white] at (8.,-1.0) {11};
\node[circle,minimum size=0.8cm,draw=black,fill=white] at (8.,-2.0) {11};
\node[circle,minimum size=0.8cm,draw=black,fill=white] at (9.,1.0) {10};
\node[circle,minimum size=0.8cm,draw=black,fill=lightgray] at (9.,0.0) {6};
\node[circle,minimum size=0.8cm,draw=black,fill=white] at (9.,-1.0) {11};
\node[circle,minimum size=0.8cm,draw=black,fill=white] at (10.,0.0) {11};
\end{tikzpicture}
\vspace{0.25cm}
\\
\(-2N_{10b},-2N_{11},+N_{3b},+N_{10c},+2N_{6}\)
\vspace{0.5cm}
\end{center}

Finally, with probability \(N_{3c}P_{3\rightarrow10}\), 4 (3c) are destroyed
and 1 (3b), 1 (10c) and 2 (6) are created by
\begin{center}
\vspace{-0.25cm}
\(P=N_{3c}P_{3\rightarrow10}\)
\nopagebreak
\\
\vspace{0.25cm}
\begin{tikzpicture}
\draw[thick][red]  (2.,0) -- (2,1);
\draw[thick][red]  (2.,0) -- (3,0);
\draw[thick][red]  (2,1) -- (3,1);
\draw[thick][red]  (3.,0) -- (3,1);
\draw[thick][red]  (9.,0) -- (9,1);
\draw[thick][red]  (8.,1) -- (9,1);
\node[circle,minimum size=0.8cm,draw=black,fill=white] at (0.,0.0) {11};
\node[circle,minimum size=0.8cm,draw=black,fill=white] at (1.,1.0) {11};
\node[circle,minimum size=0.8cm,draw=black,fill=white] at (1.,0.0) {11};
\node[circle,minimum size=0.8cm,draw=black,fill=white] at (1.,-1.0) {11};
\node[circle,minimum size=0.8cm,draw=black,fill=white] at (2.,2.0) {11};
\node[circle,minimum size=0.8cm,draw=black,fill=lightgray] at (2.,1.0) {3};
\node[circle,minimum size=0.8cm,draw=red,fill=lightgray] at (2.,0.0) {3};
\node[circle,minimum size=0.8cm,draw=black,fill=white] at (2.,-1.0) {11};
\node[circle,minimum size=0.8cm,draw=black,fill=white] at (2.,-2.0) {11};
\node[circle,minimum size=0.8cm,draw=black,fill=lightgray] at (3.,1.0) {3};
\node[circle,minimum size=0.8cm,draw=black,fill=lightgray] at (3.,0.0) {3};
\node[circle,minimum size=0.8cm,draw=black,fill=white] at (3.,-1.0) {11};
\node[circle,minimum size=0.8cm,draw=black,fill=white] at (4.,0.0) {11};
\draw [very thick] [ ->] (4.6,0.0) -- (5.4,0.0);
\node[circle,minimum size=0.8cm,draw=black,fill=white] at (6.,0.0) {11};
\node[circle,minimum size=0.8cm,draw=black,fill=white] at (7.,1.0) {11};
\node[circle,minimum size=0.8cm,draw=black,fill=white] at (7.,0.0) {11};
\node[circle,minimum size=0.8cm,draw=black,fill=white] at (7.,-1.0) {11};
\node[circle,minimum size=0.8cm,draw=black,fill=white] at (8.,2.0) {11};
\node[circle,minimum size=0.8cm,draw=black,fill=lightgray] at (8.,1.0) {6};
\node[circle,minimum size=0.8cm,draw=red,fill=white] at (8.,0.0) {10};
\node[circle,minimum size=0.8cm,draw=black,fill=white] at (8.,-1.0) {11};
\node[circle,minimum size=0.8cm,draw=black,fill=white] at (8.,-2.0) {11};
\node[circle,minimum size=0.8cm,draw=black,fill=lightgray] at (9.,1.0) {3};
\node[circle,minimum size=0.8cm,draw=black,fill=lightgray] at (9.,0.0) {6};
\node[circle,minimum size=0.8cm,draw=black,fill=white] at (9.,-1.0) {11};
\node[circle,minimum size=0.8cm,draw=black,fill=white] at (10.,0.0) {11};
\end{tikzpicture}
\\
\(-4N_{3c},+N_{3b},+N_{10c},+2N_{6}\)
\vspace{0.5cm}
\end{center}
the opposite of which happens with probability 
\begin{center}
\vspace{0.25cm}
\(P=N_{10c}P_{10\rightarrow3}\)
\nopagebreak
\\
\vspace{0.25cm}
\begin{tikzpicture}
\draw[thick][red]  (8.,0) -- (8,1);
\draw[thick][red]  (8.,0) -- (9,0);
\draw[thick][red]  (8,1) -- (9,1);
\draw[thick][red]  (9.,0) -- (9,1);
\draw[thick][red]  (3.,0) -- (3,1);
\draw[thick][red]  (2.,1) -- (3,1);
\node[circle,minimum size=0.8cm,draw=black,fill=white] at (0.,0.0) {11};
\node[circle,minimum size=0.8cm,draw=black,fill=white] at (1.,1.0) {11};
\node[circle,minimum size=0.8cm,draw=black,fill=white] at (1.,0.0) {11};
\node[circle,minimum size=0.8cm,draw=black,fill=white] at (1.,-1.0) {11};
\node[circle,minimum size=0.8cm,draw=black,fill=white] at (2.,2.0) {11};
\node[circle,minimum size=0.8cm,draw=black,fill=lightgray] at (2.,1.0) {6};
\node[circle,minimum size=0.8cm,draw=red,fill=white] at (2.,0.0) {10};
\node[circle,minimum size=0.8cm,draw=black,fill=white] at (2.,-1.0) {11};
\node[circle,minimum size=0.8cm,draw=black,fill=white] at (2.,-2.0) {11};
\node[circle,minimum size=0.8cm,draw=black,fill=lightgray] at (3.,1.0) {3};
\node[circle,minimum size=0.8cm,draw=black,fill=lightgray] at (3.,0.0) {6};
\node[circle,minimum size=0.8cm,draw=black,fill=white] at (3.,-1.0) {11};
\node[circle,minimum size=0.8cm,draw=black,fill=white] at (4.,0.0) {11};
\draw [very thick] [ ->] (4.6,0.0) -- (5.4,0.0);
\node[circle,minimum size=0.8cm,draw=black,fill=white] at (6.,0.0) {11};
\node[circle,minimum size=0.8cm,draw=black,fill=white] at (7.,1.0) {11};
\node[circle,minimum size=0.8cm,draw=black,fill=white] at (7.,0.0) {11};
\node[circle,minimum size=0.8cm,draw=black,fill=white] at (7.,-1.0) {11};
\node[circle,minimum size=0.8cm,draw=black,fill=white] at (8.,2.0) {11};
\node[circle,minimum size=0.8cm,draw=black,fill=lightgray] at (8.,1.0) {3};
\node[circle,minimum size=0.8cm,draw=red,fill=lightgray] at (8.,0.0) {3};
\node[circle,minimum size=0.8cm,draw=black,fill=white] at (8.,-1.0) {11};
\node[circle,minimum size=0.8cm,draw=black,fill=white] at (8.,-2.0) {11};
\node[circle,minimum size=0.8cm,draw=black,fill=lightgray] at (9.,1.0) {3};
\node[circle,minimum size=0.8cm,draw=black,fill=lightgray] at (9.,0.0) {3};
\node[circle,minimum size=0.8cm,draw=black,fill=white] at (9.,-1.0) {11};
\node[circle,minimum size=0.8cm,draw=black,fill=white] at (10.,0.0) {11};
\end{tikzpicture}
\nopagebreak
\\
\(-N_{10c},-N_{3b},-2N_6,+4N_{3c}\)
\end{center}


\begin{thebibliography}{99}

\bibitem{Potts52}
R. B. Potts, 
{\it Some generalised order-disorder transformations},
Proc. Cambridge Phil. Soc. {\bf 48},  106 (1952).


\bibitem{Wu82}
F. Y. Wu, 
{\it The Potts model}, 
Rev. Mod. Phys. {\bf 54}, 235 (1982).

\bibitem{Baxter82}
   R. J. Baxter, 
   \emph{Exactly solved models in statistical mechanics}, 
   1st edition (Academic Press, 1982).


\bibitem{Weaire84}
D. Weaire and N. Rivier, 
{\it Soap, cells and statistics - random patterns in two dimensions}, 
Contemp. Phys. {\bf 25}, 59 (1984).

\bibitem{Stavans93}
J. Stavans, 
{\it The theory of cellular structures},
Rep. Prog. Phys. {\bf 56}, 733 (1993).

\bibitem{Glazier90}
J. Glazier, M. Anderson and G. S. Grest, 
{\it Coarsening in the 2-dimensional soap froth and the large $Q$ Potts model - a detailed comparison},
Phil. Mag. B {\bf 62}, 615 (1990). 


\bibitem{Sokal00}
A. D. Sokal, 
{\it Chromatic polynomials, Potts models and all that},
Physica A {\bf 279}, 324 (2000).

\bibitem{Salas01}
J. Salas and A. D. Sokal, 
{\it Transfer matrices and partition-function zeros for antiferromagnetic Potts models. I. 
General theory and square-lattice chromatic polynomial}, 
J. Stat. Phys. {\bf 104}, 609 (2001).

\bibitem{Blatt96}
M Blatt, S. Wiseman, and E. Domany,
{\it Superparamagnetic Clustering of Data},
Phys. Rev. Lett. {\bf 76}, 3251 (1996).

\bibitem{Reichardt04}
J. Reichardt and S. Bornhold,
{\it Detecting fuzzy community structures in complex networks with a Potts model},
Phys. Rev. Lett. {\bf 93}, 218701 (2004).

 
 \bibitem{Ronhovde12}
 P. Ronhovde, D. Hu,  and Z. Nussinov, 
 {\it Global disorder transition in the community structure of large-q Potts systems},
 EPL  {\bf 99}, 38006 (2012).


\bibitem{Dotsenko95a}
Vik. S. Dotsenko, Vl. S.  Dotsenko, M. Picco, and P. Pujol, 
{\it Renormalization group solution for the two-dimensional random bond Potts model with broken replica symmetry},
Europhys. Lett. {\bf  32},   425 (1995).

\bibitem{Dotsenko95b}
Vl. S. Dotsenko, M. Picco, and P. Pujol, 
{\it Renormalisation group calculation of correlation functions for the 2D random bond Ising and Potts models},
 Nucl. Phys. B {\bf 455},  701 (1995).


\bibitem{KiTh88}
T. R. Kirkpatrick and D. Thirumalai, 
{\it Mean-field soft-spin Potts glass model - statics and dynamics},
Phys. Rev. B  {\bf 37},  5342 (1988). 

\bibitem{ThKi88}
 D. Thirumalai and T. R. Kirkpatrick,
{\it Mean-field Potts glass model - initial-condition effects on dynamics and properties of metastable states},
Phys. Rev. B {\bf 38},  4881 (1988). 

\bibitem{KiThWo89}
T. R. Kirkpatrick, D. Thirumalai and P. G. Wolynes,  
{\it Scaling concepts of the dynamics of viscous liquids near an ideal glassy state}, 
Phys. Rev. A {\bf 40}, 1045 (1989). 

\bibitem{Biroli11}
G. Biroli and L. Berthier, 
{\it Theoretical perspective on the glass transition and amorphous materials},
Rev. Mod. Phys.  {\bf 83},  587 (2011). 

\bibitem{KiTh15}
T. R. Kirkpatrick and D. Thirumalai, 
{\it Colloquium: Random first order transition theory concepts in biology and physics},
Rev. Mod. Phys. {\bf 87}, 183 (2015).


\bibitem{Gunton83}
J. D. Gunton, M. San Miguel and P. S. Sahni,
in Phase Transitions and Critical Phenomena vol 8, 
eds. C Domb and J L Lebowitz (New York: Academic, 1983).

\bibitem{Binder87}
K. Binder, 
{\it Theory of first order phase transitions}, 
Rep. Prog. Phys. 50, 783 (1987).

\bibitem{Oxtoby92}
D. W. Oxtoby, 
{\it Homogenoeus nucleation: theory and experiment}, 
J. Phys.: Condens. Matter 4, 7627 (1992).

\bibitem{Kelton10}
K. F. Kelton and A. L. Greer, 
{\it Nucleation in Condensed Matter}
(Elsevier, Amsterdam, 2010).  


\bibitem{Baxter73}
R. J. Baxter, 
{\it Potts model at the critical temperature},
J. Phys. C {\bf 6}, L445 (1973). 

\bibitem{Mittag74}
L. Mittag and M. J. Stephen,
{\it Mean-field theory of the many component Potts model},
J. Phys. A: Gen. Phys. {\bf  7}, L109 (1974). 

\bibitem{Baracca83} 
A. Baracca, M. Bellesi, R. Livi, R. Rechtman, and S. Ruffo,
{\it On the mean field solution of the Potts model}, 
Phys. Lett. A {\bf 99}, 156 (1983).

\bibitem{Binder81}
K. Binder, 
{\it Static and dynamic critical phenomena of the two-dimensional $q$-state Potts model},
J. Stat. Phys. {\bf 24}, 69 (1981).


\bibitem{Nam08}
K. Nam, B. Kim and S. J. Lee, 
{\it Nonequilibrium critical relaxation of the order parameter and energy in the two-dimensional ferromagnetic Potts model},
Pays. Rev. E {\bf 77}, 056104 (2008).

\bibitem{Huang10}
X. Huang, S. Gong, F. Zhong and S. Fan, 
{\it Finite-time scaling via linear driving: Application to the two-dimensional Potts model}, 
Phys. Rev. E.  {\bf 81}, 041139 (2010).

\bibitem{Li18}
C. D. Li, D. R. Tan and F. J.  Jiang,
{\it Applications of neural networks to the studies of phase transitions of two-dimensional Potts models},
Annals of Physics {\bf 391}, 312 (2018).

\bibitem{Iino19}
S. Iino, S. Morita, N. Kawashima, and A. W. Sandvik, 
{\it Detecting Signals of Weakly First-order Phase Transitions in Two-dimensional Potts Models},
J. Phys. Soc. Japan  {\bf 88}, 034006 (2019).


\bibitem{Wu97}
F. Y. Wu, 
{\it The infinite-state potts model and restricted multidimensional partitions of an integer},
Mathematical and Computer Modelling {\bf 26}, 269 (1997).

\bibitem{Johansson11}
J. Johansson and M. E. Pistol, 
{\it Microcanonical entropy of the infinite-state Potts model},
Physics Research International, 
{\bf 2011}, ID 437093 (2011).

\bibitem{bur09} R. Burioni, F. Corberi, and A. Vezzani, 
{\it Complex phase-ordering of the one-dimensional Heisenberg model with conserved order parameter},
Phys. Rev. E {\bf 79}, 041119 (2009).


\bibitem{Petri08}
A. Petri, M. Ib\'a\~nez de Berganza and V. Loreto, 
{\it Ordering dynamics in the presence of multiple phases}, 
Phil. Mag. {\bf 88}, 3931 (2008).

\bibitem{Loureiro10}
M. P. O. Loureiro, J. J. Arenzon, and L. F. Cugliandolo, 
{\it Curvature-driven coarsening in the two-dimensional Potts model}, 
Phys. Rev. E {\bf 81}, 021129 (2010).

\bibitem{Loureiro12}
M. P. O. Loureiro, J. J. Arenzon, and L. F. Cugliandolo, 
{\it Geometrical properties of the Potts model during the coarsening regime}, 
Phys. Rev. E {\bf 85}, 021135 (2012).

\bibitem{Lifshitz62}
I. M. Lifshitz, 
{\it Kinetics of Ordering During Second-Order Phase Transitions},
JETP {\bf 42}, 1354 (1962).

\bibitem{Ferrero07}
E. E. Ferrero and S. A. Cannas, 
{\it Long-term ordering kinetics of the two-dimensional q-state Potts model},
Phys. Rev.  E {\bf 76}, 031108 (2007).

\bibitem{Ibanez07b}
M. Ib\'a\~nez de Berganza, E. E. Ferrero, S. A. Cannas, V. Loreto, and A. Petri,
{\it Phase separation of the Potts model in the square lattice},
Eur. Phys. J. Special Topics {\bf 143}, 273  (2007).

\bibitem{Olejarz13}
J. Olejarz, P. Krapivsky and S. Redner, 
{\it Zero-temperature coarsening in the 2d Potts model}, 
J. Stat. Mech. P06018 (2013).

\bibitem{Denholm}
J. Denholm and S. Redner,
{\it Topology-controlled Potts coarsening},
Phys. Rev. E {\bf 99}, 062142 (2019).

\bibitem{Meunier00}
J. L. Meunier and A. Morel, 
{\it Condensation and Metastability in the 2D Potts Model}, 
Eur. Phys. J. B {\bf 13}, 341 (2000).

\bibitem{BerganzaEPL}
M. Ibáñez Berganza, P. Coletti, A. Petri,
{\it Anomalous metastability in a temperature-driven transition},
EPL {\bf 106}, 56001 (2014).

\bibitem{Ferrero09}
E. S. Loscar, E. E. Ferrero, T. S.  Grigera and S. A. Cannas,
{\it Nonequilibrium characterization of spinodal points using short time dynamics},
J. Chem. Phys. {\bf 131}, 024120 (2009).

\bibitem{Ferrero11}
E. E. Ferrero, J. P. De Francesco, N. Wolovick and S. A. Cannas,
{\it q-state Potts model metastability study using optimized GPU-based Monte Carlo algorithms},
Comp. Phys. Comm. {\bf 183}, 1578 (2011).

\bibitem{Corberi19}
F. Corberi, L. F. Cugliandolo, M. Esposito, and M. Picco, 
{\it Multinucleation in the first-order phase transition of the 2d Potts model},
 J. Phys. Conf. Series  {\bf 1226},  012009  (2019). 


\end{thebibliography}
\end{document}